\documentclass{aa}
\usepackage{graphicx}
\usepackage{subcaption}  % For side-by-side subfigures
\usepackage{caption}  
\usepackage{txfonts}
\usepackage[usenames]{color} %to change colors in references
\usepackage{tabularx} %to be more flexible placing the table

\usepackage{booktabs} 
\usepackage{footnote}
\makesavenoteenv{table*}
\usepackage{rotating}
\usepackage{float}
\usepackage{longtable} %long tables
\usepackage{lscape} %rotate table
\usepackage{pdflscape}  % or \usepackage{lscape}
\usepackage[dvipsnames]{xcolor}

\usepackage{array}
\newcolumntype{H}{>{\setbox0=\hbox\bgroup}c<{\egroup}@{}}
\usepackage{soul}
\usepackage{scalerel}
\usepackage{tikz}
\usetikzlibrary{svg.path}
\definecolor{orcidlogocol}{HTML}{A6CE39}
\tikzset{
  orcidlogo/.pic={
    \fill[orcidlogocol] svg{M256,128c0,70.7-57.3,128-128,128C57.3,256,0,198.7,0,128C0,57.3,57.3,0,128,0C198.7,0,256,57.3,256,128z};
    \fill[white] svg{M86.3,186.2H70.9V79.1h15.4v48.4V186.2z}
                 svg{M108.9,79.1h41.6c39.6,0,57,28.3,57,53.6c0,27.5-21.5,53.6-56.8,53.6h-41.8V79.1z M124.3,172.4h24.5c34.9,0,42.9-26.5,42.9-39.7c0-21.5-13.7-39.7-43.7-39.7h-23.7V172.4z}
                 svg{M88.7,56.8c0,5.5-4.5,10.1-10.1,10.1c-5.6,0-10.1-4.6-10.1-10.1c0-5.6,4.5-10.1,10.1-10.1C84.2,46.7,88.7,51.3,88.7,56.8z};
  }
}

\newcommand\orcid[1]{\href{https://orcid.org/#1}{\mbox{\scalerel*{
\begin{tikzpicture}[yscale=-1,transform shape]
\pic{orcidlogo};
\end{tikzpicture}
}{|}}} \href{#1}{#1}}
\usepackage[breaklinks=true,colorlinks, linkcolor=blue,urlcolor=blue,citecolor=blue]{hyperref}

\RequirePackage[normalem]{ulem} %DIF PREAMBLE
\RequirePackage{color}\definecolor{RED}{rgb}{1,0,0}\definecolor{BLUE}{rgb}{0,0,1} %DIF PREAMBLE
\providecommand{\DIFaddbegin}{} %DIF PREAMBLE
\providecommand{\DIFaddend}{} %DIF PREAMBLE
\providecommand{\DIFdelbegin}{} %DIF PREAMBLE
\providecommand{\DIFdelend}{} %DIF PREAMBLE
\providecommand{\DIFaddbeginFL}{} %DIF PREAMBLE
\providecommand{\DIFaddendFL}{} %DIF PREAMBLE
\providecommand{\DIFdelbeginFL}{} %DIF PREAMBLE
\providecommand{\DIFdelendFL}{} %DIF PREAMBLE
\newcommand{\DIFscaledelfig}{0.5}
\RequirePackage{settobox} %DIF PREAMBLE
\RequirePackage{letltxmacro} %DIF PREAMBLE
\newsavebox{\DIFdelgraphicsbox} %DIF PREAMBLE
\newlength{\DIFdelgraphicswidth} %DIF PREAMBLE
\newlength{\DIFdelgraphicsheight} %DIF PREAMBLE
\LetLtxMacro{\DIFOincludegraphics}{\includegraphics} %DIF PREAMBLE
\newcommand{\DIFaddincludegraphics}[2][]{{\color{blue}\fbox{\DIFOincludegraphics[#1]{#2}}}} %DIF PREAMBLE
\newcommand{\DIFdelincludegraphics}[2][]{% %DIF PREAMBLE
        \sbox{\DIFdelgraphicsbox}{\DIFOincludegraphics[#1]{#2}}% %DIF PREAMBLE
        \settoboxwidth{\DIFdelgraphicswidth}{\DIFdelgraphicsbox} %DIF PREAMBLE
        \settoboxtotalheight{\DIFdelgraphicsheight}{\DIFdelgraphicsbox} %DIF PREAMBLE
        \scalebox{\DIFscaledelfig}{% %DIF PREAMBLE
                \parbox[b]{\DIFdelgraphicswidth}{\usebox{\DIFdelgraphicsbox}\\[-\baselineskip] \rule{\DIFdelgraphicswidth}{0em}}\llap{\resizebox{\DIFdelgraphicswidth}{\DIFdelgraphicsheight}{% %DIF PREAMBLE
                                \setlength{\unitlength}{\DIFdelgraphicswidth}% %DIF PREAMBLE
                                \begin{picture}(1,1)% %DIF PREAMBLE
                                \thicklines\linethickness{2pt} %DIF PREAMBLE
                                {\color[rgb]{1,0,0}\put(0,0){\framebox(1,1){}}}% %DIF PREAMBLE
                                {\color[rgb]{1,0,0}\put(0,0){\line( 1,1){1}}}% %DIF PREAMBLE
                                {\color[rgb]{1,0,0}\put(0,1){\line(1,-1){1}}}% %DIF PREAMBLE
                                \end{picture}% %DIF PREAMBLE
                        }\hspace*{3pt}}} %DIF PREAMBLE
} %DIF PREAMBLE
\LetLtxMacro{\DIFOaddbegin}{\DIFaddbegin} %DIF PREAMBLE
\LetLtxMacro{\DIFOaddend}{\DIFaddend} %DIF PREAMBLE
\LetLtxMacro{\DIFOdelbegin}{\DIFdelbegin} %DIF PREAMBLE
\LetLtxMacro{\DIFOdelend}{\DIFdelend} %DIF PREAMBLE
\DeclareRobustCommand{\DIFaddbegin}{\DIFOaddbegin \let\includegraphics\DIFaddincludegraphics} %DIF PREAMBLE
\DeclareRobustCommand{\DIFaddend}{\DIFOaddend \let\includegraphics\DIFOincludegraphics} %DIF PREAMBLE
\DeclareRobustCommand{\DIFdelbegin}{\DIFOdelbegin \let\includegraphics\DIFdelincludegraphics} %DIF PREAMBLE
\DeclareRobustCommand{\DIFdelend}{\DIFOaddend \let\includegraphics\DIFOincludegraphics} %DIF PREAMBLE
\LetLtxMacro{\DIFOaddbeginFL}{\DIFaddbeginFL} %DIF PREAMBLE
\LetLtxMacro{\DIFOaddendFL}{\DIFaddendFL} %DIF PREAMBLE
\LetLtxMacro{\DIFOdelbeginFL}{\DIFdelbeginFL} %DIF PREAMBLE
\LetLtxMacro{\DIFOdelendFL}{\DIFdelendFL} %DIF PREAMBLE
\DeclareRobustCommand{\DIFaddbeginFL}{\DIFOaddbeginFL \let\includegraphics\DIFaddincludegraphics} %DIF PREAMBLE
\DeclareRobustCommand{\DIFaddendFL}{\DIFOaddendFL \let\includegraphics\DIFOincludegraphics} %DIF PREAMBLE
\DeclareRobustCommand{\DIFdelbeginFL}{\DIFOdelbeginFL \let\includegraphics\DIFdelincludegraphics} %DIF PREAMBLE
\DeclareRobustCommand{\DIFdelendFL}{\DIFOaddendFL \let\includegraphics\DIFOincludegraphics} %DIF PREAMBLE
\usepackage{threeparttablex} % for "ThreePartTable" environment
\usepackage{geometry}   % in preamble

\begin{document} 

   \title{Solar Orbiter observations of solar energetic electron events associated with hard microflares}

   \titlerunning{Solar Orbiter observations of SEE events associated with HMFs}

   %\subtitle{}

   \author{D.~Mittaine\inst{1,2,3} 
          \and 
          A. F. Battaglia\inst{2} 
          \and
          L. Rodríguez-García\inst{3,4}
          \and
          N. Janitzek\inst{3}
          \and  \\
          R. Gómez-Herrero\inst{4}
          \and
          F. Espinosa Lara\inst{4}
          \and
          L. Harra\inst{5,1}}
          
   \institute{ETH-Zürich, Hönggerberg campus, HIT building, Wolfgang-Pauli-Str. 27, 8093 Zürich, Switzerland
         %\\
        %\email{diane2000@orange.fr}
        \and
        Istituto ricerche solari Aldo e Cele Daccò (IRSOL), Faculty of Informatics, Università della Svizzera Italiana, Locarno, Switzerland
        \and
        European Space Agency (ESA), European Space Astronomy Centre (ESAC), Camino Bajo del Castillo s/n, 28692 Villanueva de la Cañada, Madrid, Spain
        \and Universidad de Alcalá, Space Research Group (SRG-UAH), Plaza de San Diego s/n, 28801 Alcalá de Henares, Madrid, Spain
        \and PMOD/WRC, Dorfstrasse 33, CH-7260 Davos Dorf, Switzerland
        }

   \date{Received xxx, 2026; accepted xxxxx, 2026}

%% -------------------------------------- %%
%  ABSTRACT
%% -------------------------------------- %%
 
  \abstract
  {Generally, large solar flares accelerate electrons to high energies more efficiently than microflares. However, there are microflares, known as hard microflares (HMFs), that also produce high-energy electrons, as evidenced by their flat hard X-ray (HXR) spectra. These events are characterized by a footpoint rooted in or at the edge of a sunspot. The mechanisms underlying this efficient acceleration, and their connection to solar energetic electrons (SEEs), remain to be understood.}
  {We aim to compare, for the first time, the HXR spectra of HMFs with in-situ electron spectra of associated SEEs, for which we use observations from Solar Orbiter STIX and EPD instruments, respectively. This will provide new insights into the acceleration processes and how these high-energy electrons access interplanetary space.}
   {We found eight HMFs (of GOES C class or lower), jointly observed by Solar Orbiter and Earth-bound observatories, associated with SEEs. 
   This association was verified through timing and magnetic-connectivity analysis.
   Each event was analyzed through STIX HXR spectroscopy, SEE velocity-dispersion analysis, and in-situ electron spectral analysis of EPD data.}
   {Seven out of eight SEE events show consistent timing between HXR emission from the flare and estimated electron injection into interplanetary space. They also exhibit good agreement between the flare location and the estimated magnetic connectivity to the Sun.
   The previously established correlation between HXR photon and in-situ electron spectral indices extends to HMFs, where these events populate the hard end of the distribution, even when compared to samples that include larger C- and M-class flares. This demonstrates, also through in-situ observations, that HMFs efficiently accelerate electrons to high energies, and that efficient acceleration does not necessarily require large flare energy release.
   }
   {We conclude that HMFs produce prompt SEEs with hard spectral properties, consistent with their hard HXR spectra. This indicates efficient electron acceleration to high energies. The magnetic morphology in HMFs, which involves open field lines from the sunspot, suggests that these events may be an important contributor to filling the heliosphere with energetic particles.
   }
   
   \maketitle

%% -------------------------------------- %%
%  INTRODUCTION
%% -------------------------------------- %%

\section{Introduction}
\label{sec:Introduc}
Solar flares are characterized by the rapid release of magnetic energy in the solar corona, presumably triggered when stressed magnetic field lines reconnect at the current sheet through magnetic reconnection \citep[e.g. reviews by][]{fletcher2011overview,Benz2017,arnold2021reconnection}. 
The precise mechanism underlying the efficient acceleration of particles to high energies is still debated and constitutes one of the major open questions in solar astrophysics. Upon acceleration from the corona, these particles subsequently either travel downward along magnetic field lines towards the dense chromosphere, producing emission across the electromagnetic spectrum, from radio \citep[e.g.][]{bastian1998radio,gary2018eovsa} up to hard X-rays \citep[HXRs; e.g.][]{masuda1994flare,krucker2015hxrwl} and $\gamma$-rays \citep[e.g.,][]{hurford2006gammarays,battaglia2025gammarays}, or escape outward along open field lines into interplanetary (IP) space \citep[e.g.][]{Krucker2011}.

As electrons precipitate downward, they deposit energy in the chromosphere, heating the plasma to several million Kelvin and producing thermal soft X-ray (SXR) emission \citep{Benz2017}, while nonthermal bremsstrahlung generated by collisions with the dense chromospheric plasma gives rise to HXR emission at the loop footpoints \citep{Brown1971}. Both components can be observed by the Spectrometer/Telescope for Imaging X-rays \citep[STIX;][]{Krucker2020}, the X-ray telescope aboard the Solar Orbiter mission \citep{Muller2020}, with the nonthermal HXR emission in particular demonstrating that electrons are efficiently accelerated in flares and carry a significant fraction of the released energy \citep[e.g.][]{Warmuth&Mann2020}.

Microflares are dynamic, small-scale energy-release events whose total energies are several orders of magnitude lower than those of typical M- or X-class flares. Regular microflares are generally characterized by steep (soft) HXR spectra, indicating that they are comparatively less efficient at accelerating electrons to high energies \citep[e.g.][]{Battaglia2005,Hannah2008,Warmuth&Mann2016}. However, recent observations \citep[e.g.][]{Battaglia2024} show that a small subset of microflares, known as hard microflares (HMFs), efficiently accelerate electrons to high energies. They are characterized by a particularly hard (or flat) HXR spectrum and are associated with strong magnetic fields, with their footpoints rooted in or at the edge of a sunspot \citep{Battaglia2024,Saqri2024}, making HMFs excellent case studies for investigating electron acceleration.

Solar energetic electron (SEE) events are enhancements of electron intensity detected in interplanetary space, typically spanning energies from a few keV to several hundred keV \citep[e.g.][]{Wang2012,Dresing2020}. A subset of these, known as impulsive SEE events, exhibit short rise times and strong pitch-angle anisotropies, indicating that the escaping electrons undergo only limited scattering during IP propagation \citep[e.g.][]{Krucker2009,Dresing2020,lario2024novemberevents}. Impulsive SEE events are closely associated with HXRs and type-III radio bursts \citep[e.g.][]{Lin1985,Wang2012,Warmuth2025}, with the latter reflecting electrons that escape the Sun along open magnetic field lines. 
The SEE-associated flares typically occur near the magnetic footpoints connected to the spacecraft \citep[e.g.][]{Warmuth2025}, indicating that the acceleration operates in compact regions such as active regions or individual flare sites. Therefore, SEE events constitute a powerful diagnostic tool for probing the acceleration and transport of flare-accelerated electrons. SEE events can be observed with the Energetic Particle Detector \citep[EPD;][]{Rodriguez-Pacheco2020,WimmerSchweingruber2021} onboard Solar Orbiter, which provides measurements of electron intensities, timing, and particle directivity across a broad energy range.

Given their unique properties, particularly their efficiency in accelerating high-energy electrons, HMFs are promising candidates for advancing our understanding of the mechanisms responsible for the acceleration and release of associated impulsive SEEs. This naturally motivates a comparison between the energy spectra of the flare-accelerated electrons inferred from HXR observations and those measured in-situ as escaping SEEs, this being the main objective of this study. 

A direct comparison between remote-sensing and in-situ electron spectra is however not straightforward, primarily due to IP transport effects. Nevertheless, previous studies have shown that robust correlations between them can be identified. Notably, \citet{Krucker2007} reported strong linear correlations (with coefficient $\sim$0.8) between HXR photon indices measured by the Reuven Ramaty High-Energy Solar Spectroscopic Imager \citep[RHESSI;][]{Lin2002_RHESSI} and in-situ electron spectra measured by the Energetic Particle Investigation \citep[3DP;][]{1995Lin3DPWind} on board the Wind spacecraft \citep{1997Ogilvie} for prompt events, which are the events whose inferred injection times at the Sun coincide with the HXR burst. For delayed events (inferred injection delayed by more than $\gtrsim$10 min), the correlation is weaker. 
Likewise, \citet{Dresing2021} found significant linear correlations (also with coefficient $\sim$0.8) between HXR spectral indices using RHESSI and in-situ electron spectral indices, using the Solar Electron and Proton Telescope \citep[SEPT,][]{Mueller2008SEPT} on board the Solar TErrestrial RElations Observatory \citep[STEREO;][]{Kaiser2008STEREO}. The correlation is particularly evident for events exhibiting strong anisotropy, highlighting that transport processes can modify the spectral imprint of the injected electrons. Interestingly, most events in \citet{Dresing2021} fall into the delayed category, as defined by \citet{Krucker2007}, also suggesting that some reported delays may be largely instrumental rather than physical. These studies interpret their results within the classical thin- and thick-target HXR models, in which the injected electron spectral index $\delta$ relates to the photon index $\gamma$ according to $\delta=\gamma_{thick}+1$ or $\delta=\gamma_{thin}-1$, respectively \citep{Brown1971,Holman2011}. In this work, we build upon the foundations established by both \citet{Krucker2007} and \citet{Dresing2021}, and the spectral correlations derived from our HMFs will be directly compared with those reported in the aforementioned studies.

This article is organized as follows. In Sect. \ref{sec:instr_event_selection} we present the instrumentation, the HMFs selection process, and their associated SEEs. We outline in Sect. \ref{sec:data_analysis} the procedures used to analyse the in-situ electron measurements from EPD, as well as the spectroscopic and imaging analysis of the STIX HXR observations. The results of the analysis are presented in Sect.~\ref{sec:res}, where the HXR and in situ spectra are compared. In Sect.~\ref{sec:discussion}, we discuss the interpretation of our findings. %, including spectral morphology, transport effects, and their relation to previous studies. 
Our conclusions are summarised in Sect.~\ref{sec:Conclusions}.

%% -------------------------------------- %%
%  INSTRUMENTATION AND EVENT SELECTION
%% -------------------------------------- %%
%%%BEGIN TABLE LIST OF EVENTS%%%
\begin{table*}[t]
  \centering
  \small
  %\begin{ThreePartTable}
   \setlength{\tabcolsep}{3pt}

    \caption{SEE events measured by Solar Orbiter associated to HMFs detected by STIX.}
    \label{Table:HMF-SEP_list}

    % Tabularx makes it easy to fit to the full textwidth
    \begin{tabularx}{\textwidth}{@{} cccc cc cccc @{}}
      \toprule
      \multicolumn{4}{c}{HMF} & \multicolumn{6}{c}{SEE event} \\
      \cmidrule(lr){1-4}\cmidrule(lr){5-10}
      Date & HXR peak time & Flare & $\gamma$&
      $R$ & CA & $\delta_{1}$ & $\delta_{2}$ & $\delta_{3}$ & $E_{b}$ \\
      \hline
      (UT) &  & loc [class] &  & (au) & (deg) & (-) & (-) & (-) & (-) \\ (1)& (2)&(3)&(4)&(5)&(6)&(7)&(8)&(9)&(10)\\
      \midrule

    2021/05/09 & 13:45:59 & N16E052 [C4.0] & $2.44\pm0.10$ & 0.92 & -1  & $-2.36\pm0.54^*$ & $-4.15\pm0.18^+$& - & $102\pm14$ \\
    
    2021/05/23 & 04:26:32 & N21E015 [B5.0] & $3.50\pm0.14$ & 0.95 & -5  & $-1.14\pm0.40^*$ & $-4.29\pm0.42^+$& - & $74\pm9$\\
    
    2022/02/03 & 20:33:52 & N15W064 [C1.0] & $3.64\pm0.08$ & 0.83 & 4  & $-2.09\pm0.82$ & $-2.78\pm0.10^{*+}$& - & $19\pm6$\\
    
    2022/11/10 & 17:06:28 & N13W002 [B9.0] & $3.11\pm0.13$ & 0.61 & 11  & $-2.42\pm0.50$ & $-3.87\pm0.43^{*+}$& - & $65\pm13$ \\
    
    2022/11/12 & 04:24:42 & N11W020 [B8.0] & $2.75\pm0.04$ & 0.63 & 9  & $-2.97\pm0.10^{*+}$ & - & - & - \\
    
    2022/12/23 & 04:26:28 & N17W004 [C2.0] & $2.73\pm0.06$ & 0.92 & -1 &  $-2.91\pm0.32^{*}$ & - & - & - \\
    
    2022/12/24 & 04:02:51 & N16W017 [C4.0] & $3.47\pm0.06$ & 0.93 & -3 &  $-0.53\pm0.13$ & $-1.62\pm0.05^{*+}$& $-7.13\pm4.68$ &\begin{tabular}{cc} $23\pm2$ , $284\pm68$ \end{tabular}\\
    
    2024/11/15 & 20:21:57 & S09W031 [A8.0] & $2.43\pm0.06$ & 0.78 & -15 & $-1.27\pm0.39^{*}$ & $-2.25\pm0.36^{+}$& - & $76\pm20$\\
      \bottomrule
        \end{tabularx}
    
      %\end{ThreePartTable}
\footnotesize{\textbf{Notes.} Col. 1--2: STIX HXR peak time at the Sun. Col. 3: Stonyhurst coordinates and class (in squared brackets) of the flare. Col. 4: HXR photon spectral index. Col. 5: Radial distance from the Sun to Solar Orbiter. Col. 6: Connectivity angle. Cols. 7--9: SEE spectral indices. Col. 10: Energy break. The superscript $^*$ indicates that the spectral index also corresponds to $\delta_{70}$,
$^+$ to $\delta_{200}$}, and
$^{*+}$ to $\delta_{70}=\delta_{200}$.% Details given in the main text. 
\end{table*}
%%%END TABLE LIST OF EVENTS%%%%

\section{Instrumentation used and event selection} \label{sec:instr_event_selection}
\subsection{Instrumentation}\label{sec:instrumentation}

Our study combines observational data collected from both remote-sensing and in-situ instruments onboard Solar Orbiter, particularly focusing on STIX and EPD. HMFs were studied using STIX, which observes HXR emission in the 4–150 keV range through combined imaging and spectroscopic measurements. These observations provide key diagnostics of flare-accelerated electrons and heated plasma at temperatures exceeding  $\sim8$ MK \citep{Krucker2020,Battaglia2021stix}.

SEE events associated with HMFs were directly identified and characterised in-situ using EPD, which is specifically designed to investigate suprathermal and energetic particles in the heliosphere \citep[e.g.][]{2021Gomez-Herrero}. The instrument measures electrons over energies ranging from a few keV to the near-relativistic regime (several MeV), as well as ions up to 500 MeV nuc$^{-1}$. In this study, we primarily used the Electron Proton Telescope (EPT), which covers electron energies from 25 to 475 keV. EPT consists of two double-ended telescopes and is integrated with the High Energy Telescope (HET) into two combined EPT–HET units that provide directional information on the incoming particles. EPT–HET~1 points sunward (Sun) and anti-sunward (anti-Sun) along the nominal Parker spiral, while EPT–HET~2 points northward (North) and southward (South) \citep[as shown in Fig.~4 by][]{Rodriguez-Pacheco2020}. 

We also used the SupraThermal Electrons and Protons (STEP) instrument, which measures electrons in the 2–80 keV range (effectively $\sim$4–60~keV after calibration). The STEP sensor uses two co-aligned sensor heads with the same field of view (FOV), which points in the direction of the average nominal Parker spiral. One sensor includes a permanent magnet that deflects electrons and therefore measures only ions (the magnet channel, MC), while the other measures both electrons and ions (the integral channel, IC). The electron count rate in the suprathermal energy range is obtained from the difference between the count rates of the two channels. 

\subsection{Event selection}\label{sec:event_selection}
For this study, we adopted a generic observational definition to classify an event as a HMF, mainly following characteristics reported by \citet{Battaglia2024}. In this context, HMFs are identified based on four main observational properties: relatively flat spectrum through HXR spectroscopy, a low flare intensity corresponding to GOES C-class or below, one of the flare footpoints rooted in or at the edge of a sunspot, and visible from Earth. We required visibility from Earth because it allows clear identification of whether the flare is rooted in or at the edge of a sunspot.

An initial set of HMF candidates was retrieved from the list of 39 events of \citet{Battaglia2024}. From this list, four events are potential candidates to be associated with SEEs: 3 February 2022, 10 and 12 November 2022, and 23 December 2022. These events were then cross-referenced with the Comprehensive Solar Energetic Electron event Catalogue (CoSEE-Cat) from \citet{Warmuth2025}, which includes key parameters of SEE events along with their associated flares, coronal mass ejections (CMEs), and radio bursts. By consulting the CoSEE-Cat with a focus on small-scale impulsive events, independently of the \citet{Battaglia2024} list, we identified three additional HMFs with associated SEE events happening on 9 and 23 May 2021 and on 24 December 2022. An additional HMF, occurring on 15 November 2024, was identified outside both aforementioned lists. Therefore, we identified eight HMFs potentially related to SEE events.

For each SEE candidate, we first evaluated whether the corresponding HMF could represent the solar source of the SEE by assessing its magnetic connectivity to Solar Orbiter. This was done using the Magnetic Connectivity Tool~\citep{Rouillard2020}, considering well-connected events when the heliographic Carrington longitude difference between the STIX flare location and the magnetic footpoint connected to Solar Orbiter was within $\pm20^{\circ}$, consistent with the criterion used in~\citet{Warmuth2025}. The connectivity results were further cross-checked with the Flare–SEP Linkage Quicklook Tool~\citep{SO-Flink}, which automatically links STIX flares to SEP events detected by EPD.

The timing correspondence between each HMF and its associated SEE was then evaluated using the inferred injection times obtained from the velocity dispersion analysis method \citep[VDA; e.g.][]{Vainio2013}, as described in detail in Sect. \ref{sec:data_analysis}. The injection time is expected to coincide with the peak of the nonthermal HXR emission or within the duration of the nonthermal HXR emission. All STIX times have been expressed as HXR times at the Sun. Table~\ref{Table:HMF-SEP_list} shows in Cols. 1--2 the date and HXR peak time of the selected HMFs, respectively, and the radial distance and connectivity angle (CA) of Solar Orbiter at the time of the SEE events are shown in Cols. 5--6, respectively. The CA is defined as the longitudinal separation between the flare location and the magnetic footpoint of Solar Orbiter, as provided by the Magnetic Connectivity Tool.

%----------
  \section{Data analysis}\label{sec:data_analysis}
The HXR emission for each HMFs was analysed using STIX data. The analysis relied on both spectral and imaging observations to characterize the nonthermal emission and the spatial structure of the HXR sources. 
To characterize each SEE event, key parameters were derived from EPD observations, including the onset times of the electron intensity profiles and the corresponding solar release times, which are also referred to as SEE injection times (into IP space). We also constructed the in-situ peak flux energy spectra of the solar energetic electrons measured by EPD for each SEE event \citep[similar to e.g.][]{Dresing2020, Rodriguez-Garcia2023a}, using the SERPENTINE software \citep{Palmroos2025}. We detail both analyses in the following subsections.

    %----------
%%%%BEGIN FIGURE STIX IMAGES%%%%
    \begin{figure*}
        \centering
        \includegraphics[width=\linewidth]{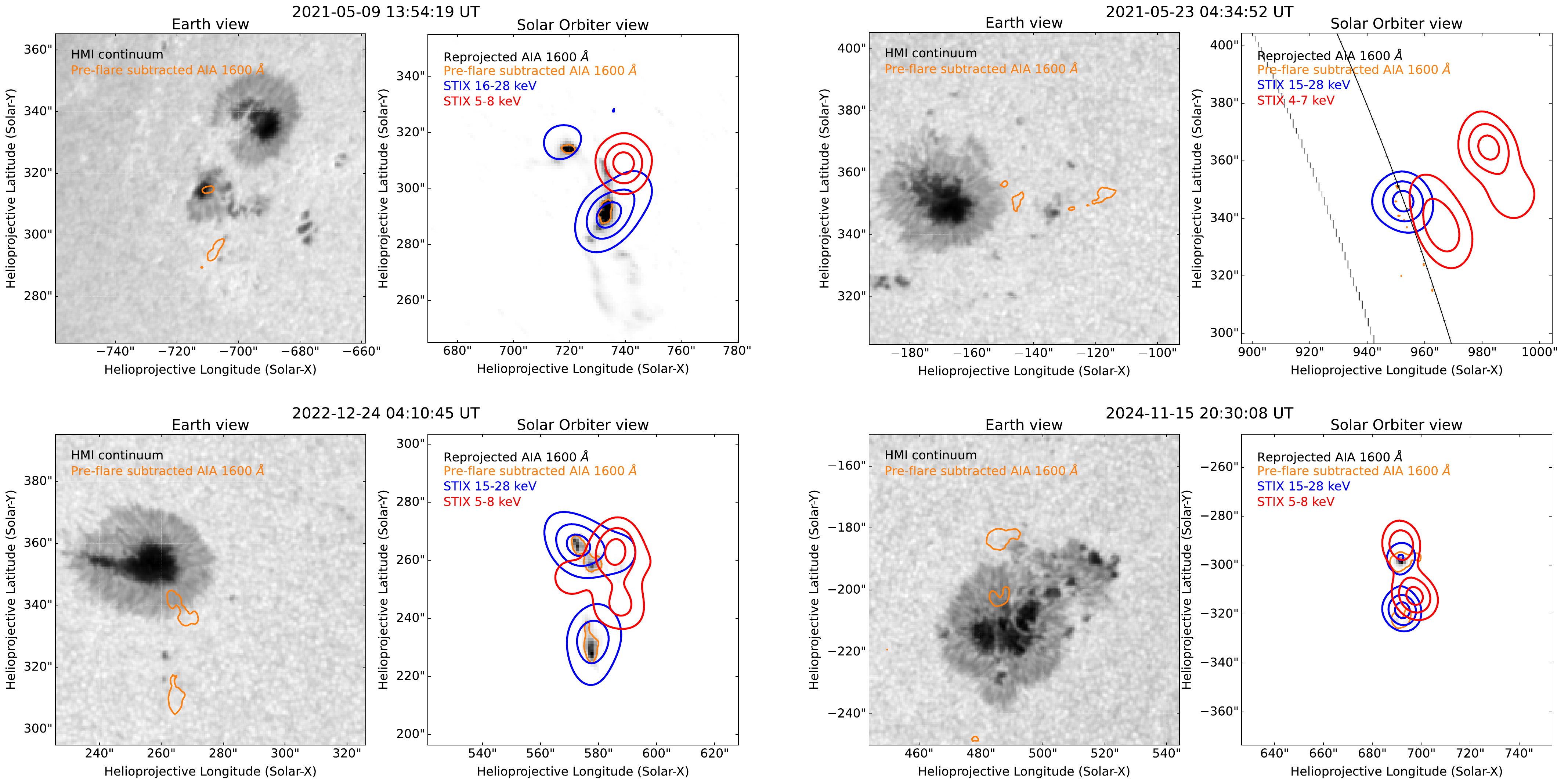}
        \caption{Solar Orbiter/STIX images of the four hard microflares not present in \citet{Battaglia2024} and representing flares in lines 1, 2, 7, and 8 of Table \ref{Table:HMF-SEP_list}. For each event, the left panel displays the SDO/HMI intensity image and the pre-flare subtracted SDO/AIA 1600 \AA{} contours (orange) from the Earth’s perspective. The right panel shows the SDO/AIA 1600 \AA{} image reprojected to the Solar Orbiter view, with orange the reprojected pre-flare subtracted SDO/AIA 1600 \AA{} contours. The STIX images are displayed as red and blue contours for low and high-energies, respectively.}
        \label{fig:overview-figure}
    \end{figure*}
%%%END FIGURE STIX IMAGES%%%%

\subsection{Analysis of the HXR observations}
\label{sub:data-analysis-HXR}
Spectroscopic data from STIX captured the energy distribution of the flare-emitted HXRs, which is essential for deriving the nonthermal spectral index. The STIX spectrogram data \citep{Krucker2020} were analysed using the OSPEX package \citep{Schwartz2002} in SolarSoftWare \citep{Freeland1998}, following the methodology described in \citet{Battaglia2024}. Spectra were fitted around the high-energy peak, with integration times of about 10–20 seconds depending on the event statistics. We note that to account for the varying distance between Solar Orbiter and the Sun compared to the Earth, all time measurements made by STIX were adjusted and expressed at the Sun.
To obtain the background-subtracted spectra, we subtracted STIX observations during quiet periods closest to each flare. The low-energy part of the spectra was fitted with an isothermal component using the \texttt{vth} function in OSPEX. 
At higher energies, the nonthermal emission was modelled with a single power law using the \texttt{pow\_fit} function in OSPEX. This power-law can be approximated as
$I(\varepsilon) \sim \varepsilon^{-\gamma}$, where $\varepsilon$ is the photon energy in keV and $\gamma$ is the photon spectral index, which is related to the underlying electron spectral index $\delta$ under the thin- and thick-target assumptions ($\gamma=\delta-1$ and $\gamma=\delta+1$, respectively; \citealt{Brown1971,Holman2011}). From the OSPEX power-law fits, we therefore obtain the photon spectral index of the emitted HXRs, $\gamma$, which is reported in Col.~4 of Table~\ref{Table:HMF-SEP_list}. In addition to the direct HXR emission component, the albedo component was added to our STIX spectra to account for the HXRs reflected from the solar surface\footnote{In the study by \citet{Battaglia2024}, the albedo component was not included. However, the difference in the spectral slope remains within one or two $\sigma$ uncertainty, thus not affecting the final results.}.

Similarly to the HXR spectroscopy analysis, the reconstruction of STIX images was performed around the high-energy peak of each HMF, following the approach described in \citet{Battaglia2024}. Since HMFs typically exhibit nonthermal emission in the 10–20~keV range, where the STIX background is lowest, this energy interval provides favourable conditions for reconstructing nonthermal sources. Image reconstruction was carried out using the CLEAN algorithm \citep{Hogbom1974} with natural weighting, adopting a CLEAN beam corresponding to the angular resolution of sub-collimator~3 \citep{Krucker2020}. 
The SDO/AIA images were reprojected to the Solar Orbiter viewpoint using standard SunPy routines, and the nonthermal HXR sources were aligned with the bright flare ribbons visible in the UV emission, following the procedure of \citet{Battaglia2024}. The four events that are not in the list of \citet{Battaglia2024} are shown in Fig.~\ref{fig:overview-figure}, illustrating that also these HMFs are rooted in or at the edge of sunspots.

\subsection{Analysis of the solar energetic electrons}\label{is-data}

\subsubsection{Particle timing}
The timing of energetic particle observations is a key factor in establishing the connection between SEE events and their solar sources.
We analysed the timing of each SEE event using the VDA method \citep[e.g.][]{Vainio2013}, which assumes that electrons of different energies are released simultaneously at the Sun and that the first-arriving particles propagate scatter-free along a single effective path length~$L$, with constant speed. As these particles travel through IP space along the Parker spiral, their arrival times at Solar Orbiter depend on their velocities and, therefore, on their kinetic energies. Under these assumptions, the apparent injection time at the Sun and the corresponding path length were obtained by applying a linear fit of the measured onset times as a function of the inverse particle speed. 
The fit was done by using the scipy.odr (hereafter ODR) package in Python.

In order to determine these onset times, we used the SERPENTINE tool \citep{Palmroos2025}, which applies the statistical Poisson-CUSUM method to automatically detect the onset in each energy channel as the moment when the measured intensities start to deviate significantly from the pre-event background level. In our analysis, we defined this threshold as the mean background level plus three standard deviations ($3\sigma$). We include more details about the VDA method and the use of the SERPENTINE tool in Appendix \ref{app:appendixA1}.

For each event, we selected the EPT telescope that detected the earliest arriving electrons among the Sun, anti-Sun, North, and South directions; in most cases, this was the Sun-pointing telescope. The method used for this selection is detailed in Appendix \ref{app:appendixA1}. For weaker events with low EPT–Sun counting statistics, we considered STEP data. In our sample, only the 3 February 2022 SEE event showed low counts in EPT–Sun. Although STEP data were examined, the highest channels were too noisy with strong pre-event fluctuations, preventing reliable onset determination. Consequently, all VDA fits in this study were performed using EPT-only data. 
The number of energy channels included in the linear fit corresponds to those with a clear onset, exceeding the mean background by more than $3\sigma$. A 1-minute time averaging was used as a baseline for all events. For slow-rising or low-intensity events, longer averages (up to two minutes) were required, whereas for rapidly rising events and more generally, when the counting statistics allowed, shorter averaging (e.g. 30 seconds) was applied.

%%%%BEGIN FIGURE
 \begin{figure}
    %\centering
        \centering
        \includegraphics[width=\linewidth]{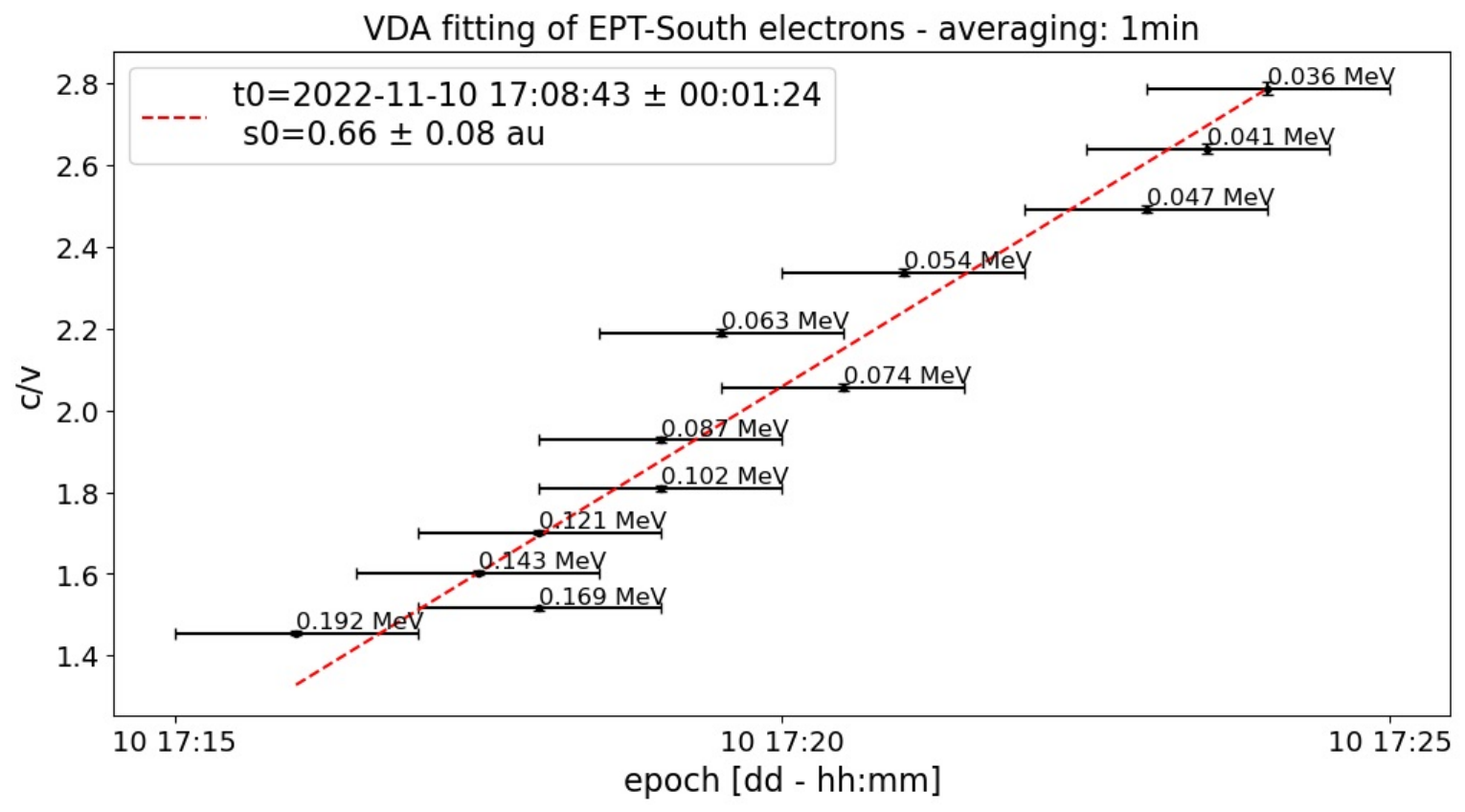}
        \caption{VDA fitting of the EPT-South electrons for the event on 10 November 2022. The vertical and horizontal axes correspond with the reciprocal of the particle velocities (with v in au/day) and onset times, respectively. The black points identify the electron onsets at the corresponding velocities (energies), with the respective errors indicated. The red dashed line is the linear regression fit to all points. The legend gives the effective path length (s0) and the estimated release time (t0). }
        \label{VDA_ex}
    \end{figure}
%%END FIGURE

As an example, Fig.~\ref{VDA_ex} illustrates the VDA fitting for the event on 10~November~2022, performed using the EPT–South telescope and a one minute averaging. For this case, energy channels between 31.2~keV and 184.9~keV were included in the fit. We note that a re-binning factor of two was applied to the data prior to fitting, to reduce statistical fluctuations. The results of the VDA fitting for all analysed events are summarized in Cols. 7--8 of  Table~\ref{Table:TSA_VDA}, presenting respectively the estimated released time and the effective path length followed by the particles. 

The time-shift analysis method (TSA) was also used to estimate the injection time of the energetic electrons, when a reliable VDA fit could not be obtained, namely in the SEE event on 24 December 2022. It assumes scatter-free propagation along a nominal Parker spiral $L_{n}$ and derives the solar injection time for a given energy $E$ from the particle onset at the spacecraft, shifted by the travel time along the magnetic field line \citep[e.g.][]{Vainio2013}. The nominal Parker spiral length ($L_{n}$) was calculated using the solar wind speed measured by the Solar Wind Analyzer \citep[SWA;][]{Owen2020}, onboard Solar Orbiter, at the time of the SEE onset. In the absence of SWA measurements, a nominal speed of 400 km s$^{-1}$ was assumed, as done in previous studies \citep[e.g.][]{Paassilta2018, Rodriguez-Garcia2023a, Warmuth2025}. The results of TSA, along with the nominal Parker spiral for each event are given in Cols. 4--6 of Table \ref{Table:TSA_VDA}. We notice that in cases where both VDA and TSA yielded reliable injection times, VDA results were used, as it provides a more direct constraint on the IP transport and is generally considered more robust when a consistent energy–dependent onset is observed.

\subsubsection{Particle spectra} \label{subsubsec:particle-spectra}
To investigate the energy distribution of the in-situ solar energetic electrons, the peak-intensity spectrum for each event was constructed. It consisted in identifying, for each energy channel and within a defined time interval, the maximum flux reached during the SEE event. This was done using the SERPENTINE tool \citep{Palmroos2025}, which automatically identifies the peak flux as the maximum mean of the resampled electron intensity within this interval, as detailed in Appendix \ref{app:appendixA1}. A time averaging of one minute was used as the baseline for this resampling. For lower-intensity or slowly varying events, a longer averaging (i.e., two to five minutes) was applied to improve the counting statistics and ensure a robust peak determination. Similarly to what has been done for VDA and when counting statistics allowed, a shorter averaging (i.e., 30 seconds) was applied. In constructing the peak-intensity spectra, an average pre-event background level was also calculated over a quiet-time interval and subtracted from the peak fluxes in each energy channel.

Our analysis focused primarily on energies within the EPT range, which provides measurements in the four distinct viewing directions. The telescope viewing direction exhibiting the highest peak intensity was selected. Whenever an event was observed in the Sun direction, the EPT observations could be combined with STEP measurements, as STEP observes in the same direction as EPT-Sun. We note that an intensity offset can occur between EPT-Sun and STEP measurements due to their different FOVs. To ensure spectral consistency, only STEP pixels within the EPT-Sun FOV were used, following the approach described by \citet{fedeli2026}.
To automatically decide whether STEP data could be combined with EPT–Sun observations for a given event, we implemented a statistical detection method based on background-level enhancement. For each event, the pixel-weighted STEP flux (restricted to pixels within the EPT–Sun FOV) was computed, and a quiet-time background interval preceding the event was selected to derive the mean background level and its standard deviation. Energy channels showing at least three consecutive points above the mean background plus $3\sigma$ were flagged as significant. Events with strictly more than five such channels ($N>5$) were considered suitable for combination with EPT–Sun data. This threshold ensures a balance between sensitivity and robustness against noise, consistent with criteria used in previous studies \citep[e.g.][]{Dresing2020,Warmuth2025}.

Data from the magnetometer instrument onboard Solar Orbiter \citep[MAG;][]{Horbury2020} for both STEP and EPT observed SEE events were also examined to assess the magnetic field direction at the time corresponding to the peak flux for each energy channel. The magnetic field was averaged using the same time cadence as applied to the electron flux.
This allowed us to identify cases where the field orientation indicated that the electron beam was directed toward outer STEP pixels or outer EPT FOV, suggesting that the corresponding energies may not represent the main beam and could therefore potentially affect the resulting spectra. More details on this cross-check with data from MAG instrument are provided in Appendix \ref{app:appendixA3}.

To fit the spectral data, we used the ODR package in Python. It accounts for uncertainties in both axes: the energy uncertainties were taken as the widths of the energy bins \citep[c.f.,][]{Dresing2020}, while the intensity uncertainties were obtained by propagating the statistical uncertainties from the resampling procedure of the SERPENTINE tool, used to derive the peak fluxes. Since the fits were performed on background-subtracted spectra, the final intensity uncertainties were computed as the quadrature sum of the peak and pre-event background uncertainties. We note that only energy channels showing a significant enhancement, at least the 3-$\sigma$ level above the pre-event background, were included in the fitting of the background-subtracted peak flux spectra.

Following previous studies \citep[e.g.,][]{Rodriguez-Garcia2023a,fedeli2026} three different fitting models were tested for fitting the background-subtracted spectra:

\begin{equation}
I(E) = I_{0}\left(\frac{E}{E_{0}}\right)^{\delta_{1}},
\label{eq:spl}
\end{equation}

 \begin{equation}
I(E) = I_{0}\left(\frac{E}{E_{0}}\right)^{\delta_{1}}\left(\frac{E^{\alpha} + E^{\alpha}_{b}}{E^{\alpha}_{0} + E^{\alpha}_{b}}\right)^{\frac{\delta_{2} - \delta_{1}}{\alpha}}, \quad \mathrm{and}
\label{eq:bpl}
\end{equation}

 \begin{equation}
I(E) = I_{0}\left(\frac{E}{E_{0}}\right)^{\delta_{1}}\left(\frac{E^{\alpha} + E^{\alpha}_{bl}}{E^{\alpha}_{0} + E^{\alpha}_{bl}}\right)^{\frac{\delta_{2} - \delta_{1}}{\alpha}} \left(\frac{E^{\beta} + E^{\beta}_{bh}}{E^{\beta}_{0} + E^{\beta}_{bh}}\right)^{\frac{\delta_{3} - \delta_{2}}{\beta}}.
\label{eq:double-bpl}
\end{equation}

\noindent
Equation (\ref{eq:spl}) corresponds to a single power-law, where $I_0$ is the intensity at the reference energy $E_0 = 0.1$ MeV, which is reference energy value chosen for this study, and $\delta_{1}$ is the spectral index. Equation (\ref{eq:bpl}) is a broken power-law, that transitions from spectral index $\delta_{1}$ to spectral index $\delta_{2}$, at  an energy $E_{b}$ (spectral break) and equation (\ref{eq:double-bpl}) is a double broken power-law, that transitions first at energy $E_{bl}$ from spectral indices $\delta_{1}$ to $\delta_{2}$ and then again, at energy $E_{bh}$ from spectral indices $\delta_{2}$ to $\delta_{3}$. Notice that $\alpha$ and $\beta$ parameters describe the sharpness of the spectral transition and all these parameters are fitted simultaneously.
The most suitable model was then selected based on both the observed spectral shape and the residual variance returned by the ODR fit, corresponding to the reduced $\chi^{2}$ value divided by the number of degrees of freedom (DOF). 

%%%%BEGIN FIGURE
\begin{figure}
    %\centering
        \centering
        \includegraphics[width=\linewidth]{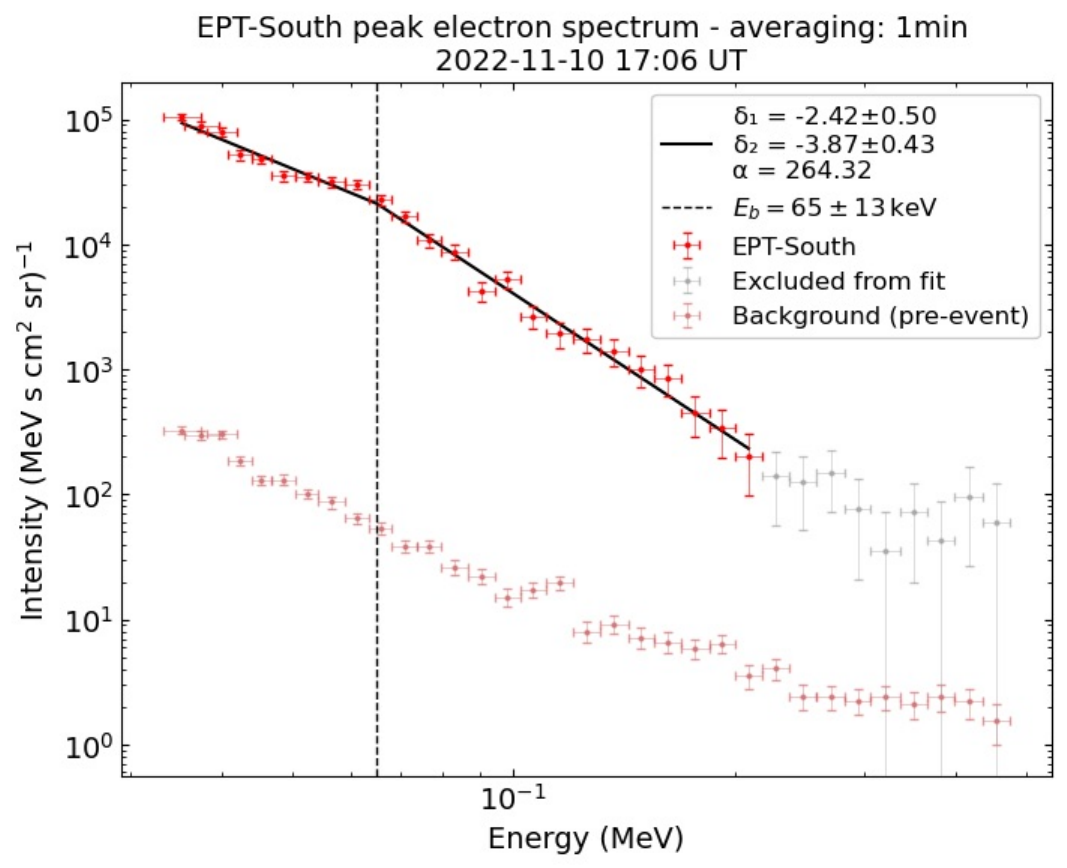}
        \caption{Background-subtracted peak flux spectrum of the EPT-South electrons for the event on 10 November 2022. The red points represent EPT-South electron peak flux. The fainter points show the pre-event background that is subtracted from the peak intensities. The gray points denote energy channels that were excluded from the fit, as they do not meet the 3-$\sigma$ significance criterion above the pre-event background. The dashed vertical black line represents the spectral break, and the legend provides the fit parameters.}
        \label{enu_ex}
    \end{figure}
%%%END FIGURE    
As an example, Fig.~\ref{enu_ex} illustrates the fitting of a broken power-law to the background-subtracted peak flux spectrum for the event on
10 November 2022, performed using the EPT–South telescope.
For this case, energy channels between 31.2 and 218.2 keV were included in the fit and the value of the residual variance is given by $0.84$, which is reasonable.
The results of the spectral fits for all analysed events are summarized in Cols. 7--10 of Table~\ref{Table:HMF-SEP_list}.

%% -------------------------------------- %%
%  RESULTS
%% -------------------------------------- %%
\section{Results}
\label{sec:res}
%----------

\subsection{Magnetic connectivity}
As shown in Col. 6 of Table \ref{Table:HMF-SEP_list}, all eight events in our sample exhibit CAs within $\pm20^\circ$. However, some limitations of the magnetic connectivity tool must be considered. In particular, for the event on 15 November 2024, the case with the largest connection angle (CA=-15$^{\circ}$), the connectivity map using the Air Force Data Assimilative Photospheric flux Transport (ADAPT) modelling framework\footnote{\url{https://nso.edu/data/nisp-data/adapt-maps/}} shows a broad spread of possible magnetic footpoints, extending over nearly $\sim20^\circ$. We selected the footpoint closest to the STIX flare location in heliographic longitude, yet this spread introduces uncertainty in the true magnetic connection. Consequently, the connectivity for this event should be interpreted with caution.
%%%%BEGIN FIGURE STIX SPECTRUM
  \begin{figure}
        \centering
        \includegraphics[width=\linewidth]{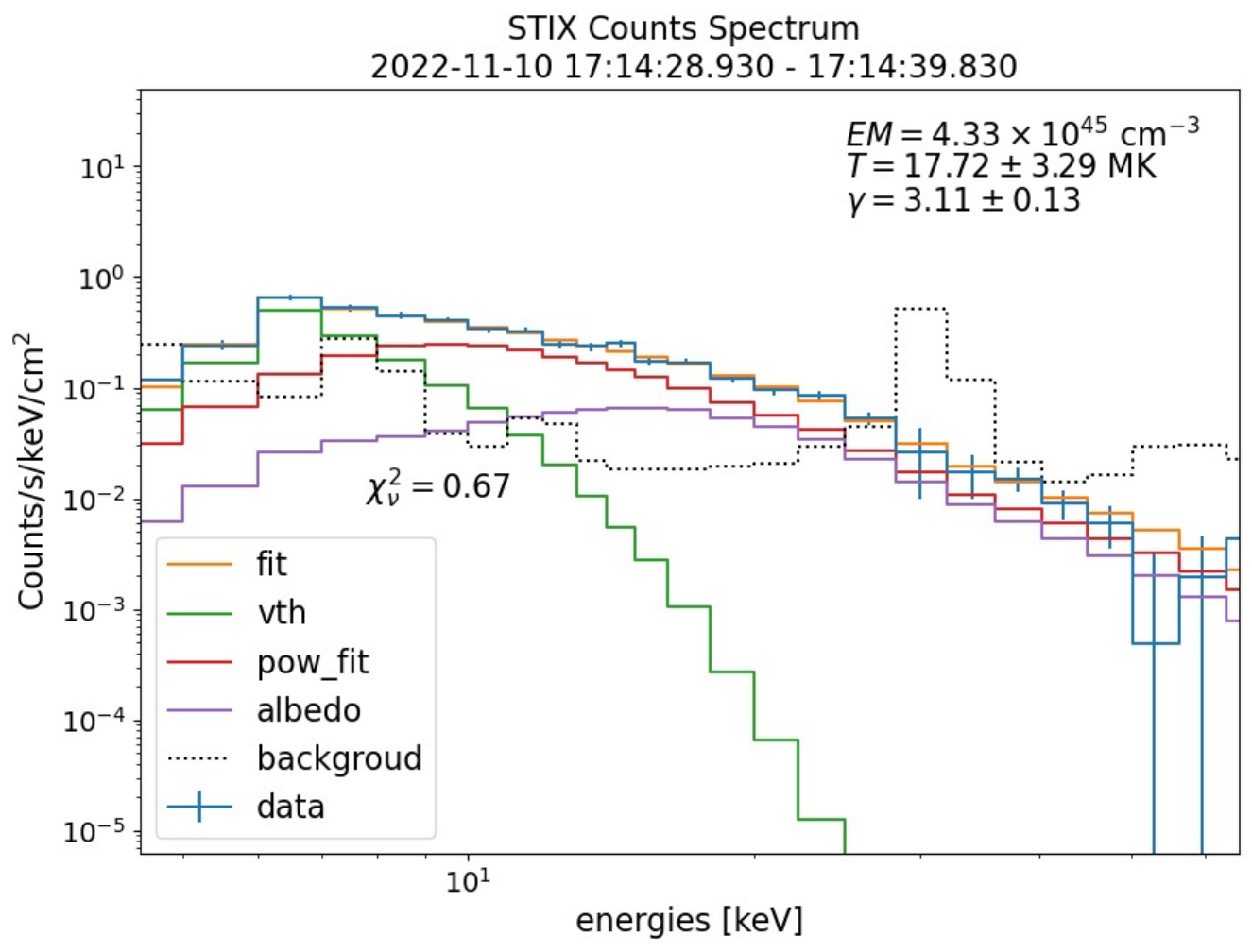}
        \caption{Pre-flare background subtracted STIX count spectrum (blue) for the HMF event on 10 November 2022. The spectrum at low energies is fitted with an isothermal fit (green curve), where $T$ and $EM$ denote the plasma temperature and the emission measure, respectively, quantifying the amount of plasma emitting at that temperature. At high energies, the spectrum is fitted with a single power-law (red curve), with $\gamma$ its photon spectral index. The albedo component (purple curve) is added to account for HXRs reflected from the solar surface. The total fitted spectrum is shown in orange, and the subtracted background in dotted black. The reduced chi-squared value $\chi^2_\nu$ of the fit is also indicated in the figure.}
        \label{fig:enu-spec-HXRs}
    \end{figure}

%%%END FIGURE STIX SPECTRUM 

\subsection{Timing}
 The quality of the fitting in terms of the corresponding residual variance was satisfactory for seven out of eight events, given low time uncertainties and physical mean free paths. However, for the 24~December~2022 SEE event, the VDA fit returned an unphysical short path length far below the nominal Parker spiral value, likely due to the response to high energy electrons in the lower energy channels \citep[c.f.][]{Warmuth2025}. Consequently, neither the VDA-inferred injection time nor path length is reported in Table \ref{Table:TSA_VDA} and we relied instead on the TSA-derived injection time, which remains consistent with the corresponding HXR peak.
 
 For the seven SEE events with feasible VDA in Table~\ref{Table:TSA_VDA}, the timing of SEE injections (Col. 8) relative to the peak times of the HXR emission (Col. 2) show time differences between 2.3 and 9.7 min, with a mean of 4.9~min and a median of 4.3~min. These delays are slightly larger than those reported for impulsive SEE events by \citet{Warmuth2025}, who found a mean delay of about 2.8~min and a median of about 3.7~min. This difference is noteworthy given that our HMFs share the characteristics of impulsive events: short, rapidly rising electron onsets associated with compact flares and brief HXR bursts. However, a direct comparison requires accounting for the timing uncertainties for both the HXR peak time and the injection time.
 
 The uncertainty of the HXR peak time is better represented by the full duration of the nonthermal HXR emission, which provides upper and lower bounds for the HXR peak. This is physically motivated by considering the fact that the electron injection onto open field lines can occur at any point during the nonthermal phase, depending on when reconnection between %closed flare loops and 
open magnetic structures takes place. 
We therefore used the start and end times of the HXR emission as asymmetric bounds on the HXR peak time, as reported in Table~\ref{Table:TSA_VDA}, and combined these with the VDA fitting uncertainties on the in-situ injection times to estimate a representative uncertainty on the delay $\Delta t = t_{\mathrm{IS}} - t_{\mathrm{HXR}}$. This yields a mean uncertainty of $\sim$4.3~min on the delay itself, comparable to the measured mean delay of 4.9~min, indicating a logical timing correspondence. Additional contributions, such as instrument response, onset determination, and the averaging involved in VDA, may further explain part of the remaining spread. 

 Finally, the path lengths obtained from the VDA fits (Col. 9 in Table~\ref{Table:TSA_VDA}) are typically within $\sim$20\% of the nominal Parker spiral lengths $L_n$ (Col. 4), except for one clear outlier on 3 February 2022 that will be discussed in Sect. \ref{sub:diss-path}. Such deviations are physically reasonable and in agreement with moderate IP scattering, supporting the reliability of the VDA solutions for the seven events where the method was applicable. 

%----------

%%%BEGIN FIGURE EPD PEAK FLUX SPECTRA
  \begin{figure*}
        \centering
        \includegraphics[width=\linewidth]{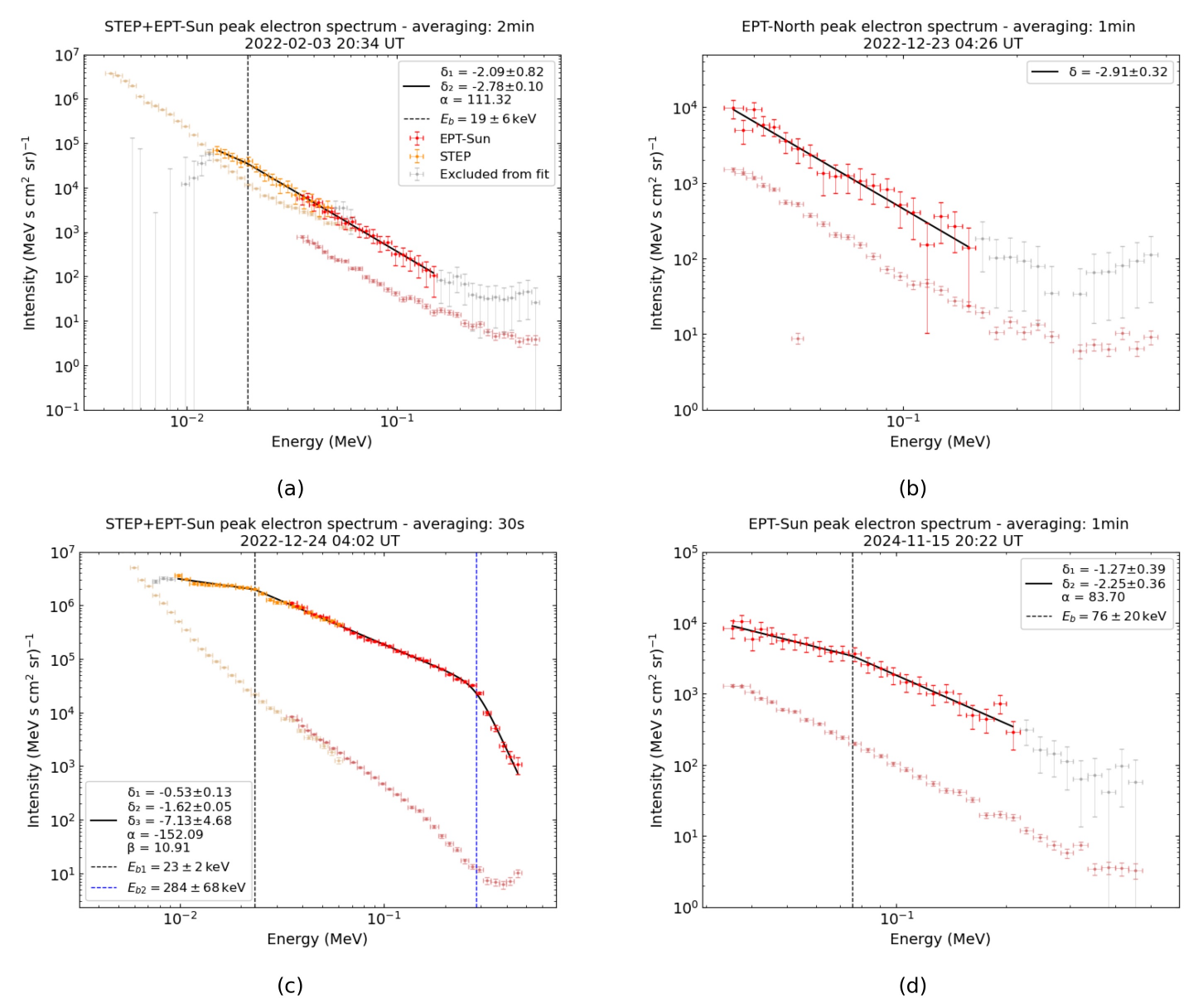}
        \caption{Background-subtracted EPD peak flux spectra for four selected events. The fainter, lower-intensity points indicate the pre-event background, which is subtracted from the peak intensities. Gray points denote energy channels excluded from the fits. (a) 3 February 2022 event, combining STEP and EPT-Sun data. Orange and red points represent STEP and EPT-Sun electrons, respectively. The dashed vertical black line marks the spectral break of the broken power-law fit. (b) 23 December 2022 event observed by EPT-North and fitted with a single power-law with spectral index $\delta$. Red points represent EPT-North electrons. (c) 24 December 2022 event, combining STEP and EPT-Sun data. Orange and red points represent STEP and EPT-Sun electrons, respectively. The dashed vertical black and blue lines indicate the two spectral breaks of the double broken power-law fit. (d) 15 November 2024 event observed by EPT-Sun. The dashed vertical black line marks the spectral break of the broken power-law fit. In all panels, the legend provides the corresponding fit parameters.}
        \label{fig:enu-spec-is}
    \end{figure*}

%%%END FIGURE EPD PEAK FLUX SPECTRA
\subsection{Spectral analysis}\label{subsec_spectral_analysis}
An example STIX spectrum, for the 10 November 2022 event, is shown in Fig.~\ref{fig:enu-spec-HXRs} (all other spectra are illustrated in Fig.~\ref{all-HXR-spectra}).

Across all analysed events, the indices range from $2.43 \pm 0.06$ to $3.64 \pm 0.14$, with a mean of $\gamma_{\mathrm{mean}} = 3.01 \pm 0.08$ and a median of $\gamma_{\mathrm{median}} = 2.93 \pm 0.08$. These values indicate relatively hard nonthermal spectra, consistent with \citet{Battaglia2024}. These events are generally harder on average than the HXR spectral indices found in earlier studies such as \citet{Krucker2007} (with mean $\gamma \sim 3.44$) and \citet{Dresing2021} (with mean $\gamma \sim 4$).

 For the in-situ electrons, we followed the procedure described in Sec.~\ref{is-data} for fitting the particle spectra. As an example, the background-subtracted peak-intensity spectra for four selected SEE events are shown in Fig.~\ref{fig:enu-spec-is}: the 3~February~2022 (a) and 15~November~2024 (d) events, for which a broken power-law fit (Eq.~\ref{eq:bpl}) is applied; the 23~December~2022 event (b), fitted with a single power law (Eq.~\ref{eq:spl}); and the 24~December~2022 event (c), for which a double broken power-law fit with two spectral transitions (Eq.~\ref{eq:double-bpl}) is required.

 The spectral parameters derived for all events are summarised in Cols. 7--10 of Table~\ref{Table:HMF-SEP_list}, while their corresponding in-situ spectra are shown in Fig.~\ref{all-is-spectra} of Appendix~\ref{additional-spec}. Across the eight analysed events, two spectra are well described by a single power law, five are best fitted by a broken power law, and one event (24~December~2022) requires a double broken power-law model, indicating two distinct spectral transitions. For the five broken power-law events, the spectra exhibit a systematic steepening from low to high energies, with a mean break energy of $E_b = 67 \pm 13$~keV. This value is consistent with the break energies reported by \citet{Krucker2009}, but is systematically lower than those found by \citet{Dresing2020}, where the mean break energy is about 120~keV. 
 In fact, simulations of electron beams propagating through the coronal and interplanetary plasma have further shown that wave–particle interactions can naturally introduce spectral breaks in the electron spectrum, with a steepening at higher energies (typically above $\sim$60~keV; \citealt{Kontar&Reid2009,Reid&Kontar2010,Reid&Kontar2013}). 
 This difference is discussed in more detail in Sect.~\ref{subsec:is-spectral-breaks}.

 Columns 7–9 of Table~\ref{Table:HMF-SEP_list} list the spectral indices derived for our event sample. Taking $\delta_{200}$ as the reference parameter for comparison with previous studies, we obtain a mean value of $\delta_{200,\mathrm{mean}} = -3.18 \pm 0.25$. This value lies toward the harder end of the distribution reported in earlier studies of SEE events near 1 au, which also included larger events. \citet{Krucker2009} investigated 62 impulsive electron events in the energy range 1–300 keV observed by the \textit{Wind} spacecraft and reported a mean value of $\langle \delta_{200} \rangle = -3.6 \pm 0.7$. Similarly, \citet{Dresing2020} analyzed 781 near-relativistic SEE events measured by both \textit{STEREO} spacecraft, finding $\langle \delta_{200} \rangle = -3.5 \pm 1.4$. We note that the events in our sample were observed at different radial distances, ranging from 0.63 to 0.95 au (see Col.~5 of Table~\ref{Table:HMF-SEP_list}). \citet{Rodriguez-Garcia2023a} reported that radial variations introduce differences in spectral hardness of approximately 10\%. Accounting for this effect would shift our mean value slightly toward softer spectra but still in the harder part of previous near 1 au studies.
 
 Figure~\ref{fig:enu-spec-is} illustrates how combining EPT–Sun and STEP observations extends the spectral coverage to lower energies (shown in panels~(a) and~(c)) and, in these cases, reveals low-energy spectral breaks that would not be detectable using EPT alone.
 We note that, although five out of eight events are observed in the EPT–Sun viewing direction, only three (3~February~2022, 12~November~2022, and 24~December~2022) fulfil the criteria for combining STEP with EPT–Sun data (shown in Sect.~\ref{subsubsec:particle-spectra} and Appendix~\ref{app:appendixA3}).  
 For the 3~February~2022 event, despite satisfying the $N > 5$ criterion defined in Sect.~\ref{is-data}, the three highest-energy STEP channels were excluded from the fit because they are noisy, exhibit strong pre-event fluctuations, and their associated magnetic-field directions at peak time point outside the STEP FOV, indicating that they do not sample the main electron beam.

 %%%BEGIN FIGURE  CORRELATION PLOT    
  \begin{figure*}
          \centering
           \includegraphics[width=\linewidth]{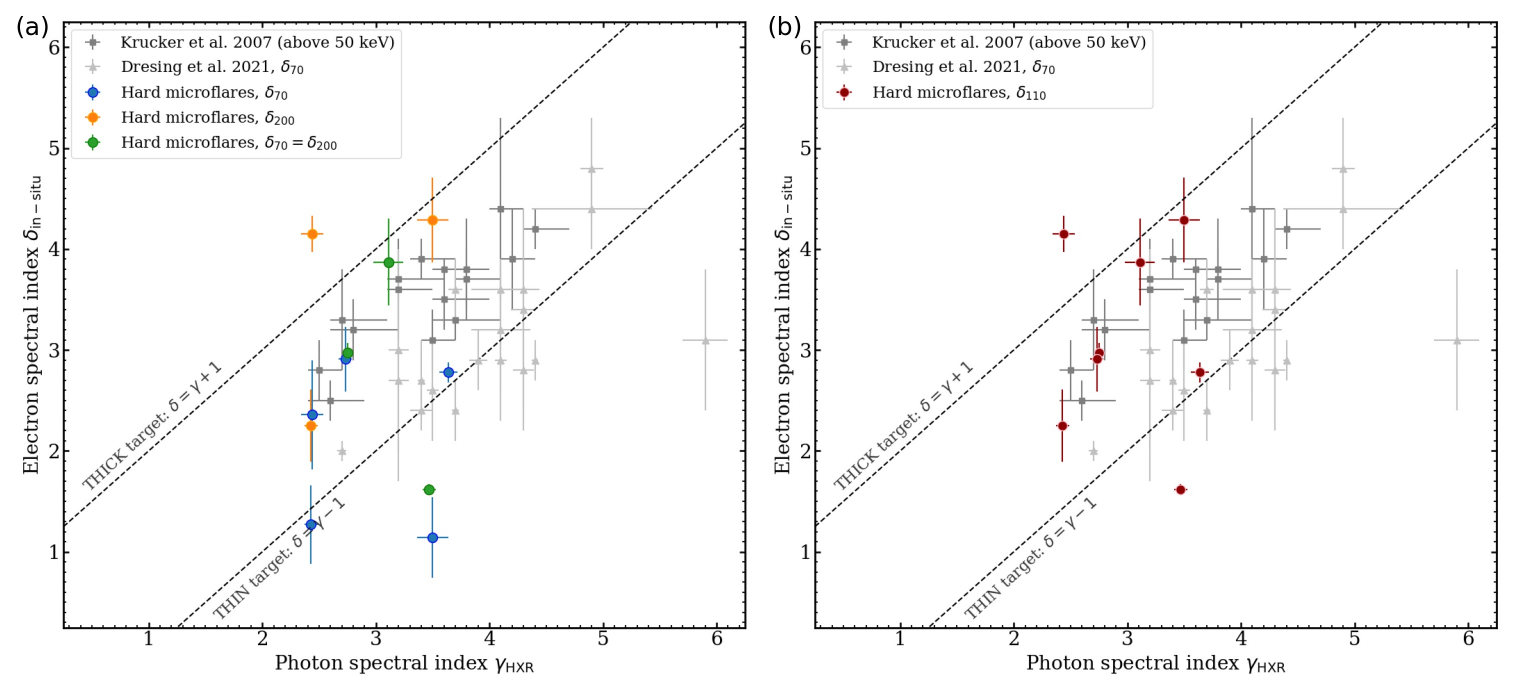}
           \caption{Correlation plots of the in-situ electron spectral index $\delta_{\mathrm{in-situ}}$ as a function of the HXR photon spectral index $\gamma_{\mathrm{HXRs}}$ at different characteristic energies: 70 and 200 keV in panel (a) and 110 keV in panel (b). The dashed gray lines represent the thick-target ($\delta = \gamma + 1$) and thin-target ($\delta = \gamma - 1$) models used to infer electron spectra from HXR observations. Previous studies are shown in gray: squares denote events from \citet{Krucker2007}, while triangles denote events from \citet{Dresing2021}. The events studied in this paper are labelled with different colours according to the energy at which the spectral index was derived as shown in the legend.}
           \label{Correlation_delta70-200}
    \end{figure*}
 %%%%END FIGURE CORRELATION PLOT
 
\subsection{Spectral relations}    
\label{res:spectral-relations}
Correlations between the spectral indices of HXR photons and the in-situ electrons have been found in previous studies \citep[e.g.][]{Krucker2007,Dresing2021}. As discussed earlier, the in-situ electron spectra may be altered by energy-dependent effects, leading to spectral breaks and, consequently, to different spectral shapes across the events. Choosing which indices to compare is therefore an important consideration. Following the approach of \citet{Dresing2020} and \citet{Rodriguez-Garcia2023a}, we first determined the in-situ spectral indices at fixed reference energies of 70~keV and 200~keV, which are indicated with markers in Table \ref{Table:HMF-SEP_list}. This enabled a comparison of the spectral indices of the different events, regardless of their spectral shapes. 

Additionally, we extended this analysis to one alternative reference energy at 110~keV. This choice is motivated by the need to distinguish between electron spectral indices that are below and above the spectral break, as there are very likely different physical processes governing the different parts of the spectrum. Here, we restricted the comparison to the in-situ electron spectral indices that are above the (first) spectral break which lies at energies up to 100\,keV in our sample. This is precisely in line with \citet{Krucker2007} who also argue that the range above the break is less likely to be affected by transport effects and thus more appropriate for analysis of the underlying electron acceleration process. 
We therefore look in a complementary analysis at in-situ spectral indices around $\sim$110~keV, as these are above the first spectral break for all of our events, and thus covered by reliable measurements in the full sample. 

 Figure~\ref{Correlation_delta70-200} shows the relationship between the HXR and in-situ electron spectral indices obtained from our study, together with the values reported by \citet{Krucker2007} and \citet{Dresing2021} as a reference. We note that in the study of \citet{Dresing2021} no $\delta_{70}$ is explicitly reported. To enable a consistent comparison with our results, we defined the reference $\delta_{70}$ from their sample. We took the corresponding spectral index at 70 keV as the one below the break when the associated break energy satisfies $E_b > 70$~keV, and the spectral index above the break, when $E_b < 70$~keV. 

 Overall, our measurements fall largely within the cloud of values that are observed by \citet{Krucker2007} and \citet{Dresing2021}.
 Interestingly, however, our sample shows on average harder spectra at both 70 and 200~keV than the comparison samples, even though we are considering small flares, as shown in Fig.~\ref{Correlation_delta70-200}a. This does not only support the classification of these events as hard microflares, but also indicates that HMFs can produce hard in-situ electron spectra, with mean spectral indices of $\delta_{70,mean}=2.36 \pm 0.29$ (median $\delta_{70,median}=2.57 \pm 0.39$) at 70~keV and $\delta_{200,mean}=3.18 \pm 0.25$ (median $\delta_{200,median}=3.41 \pm 0.25$) at 200~keV. In some cases, these values are comparable to the hardest spectra previously reported for larger events \citep[e.g.][]{Dresing2020}, as discussed in Sect. \ref{subsec_spectral_analysis}. Figure~\ref{Correlation_delta70-200}b further supports this behaviour, when extending the comparison to the alternative reference energy at 110~keV, with mean spectral index of $\delta_{110,mean}=3.10 \pm 0.25$ (median $\delta_{110,median}=2.94 \pm 0.21$). Consistently, these mean and median spectral indices for 110~keV lie between those values found for 70~keV and 200~keV.

 In general, our events align well with the sample from \citet{Krucker2007}, lying between the thin- and thick-target model predictions. However, Fig.~\ref{Correlation_delta70-200}a shows that at 70~keV some of our points tend to lie closer to the thin-target model, whereas at 200~keV a few events appear closer to the thick-target model. Note that a similar tendency is also reported in \citet{Dresing2021}, where the in-situ spectral indices associated with energies below the spectral break ($\delta_1$) align along the thin-target model, while the spectral indices associated with energies above the break ($\delta_2$), and therefore higher energies, are shifted more toward the thick-target line. 
 However, from imaging of the HXR sources in Fig.~\ref{fig:overview-figure}, we find that these are located at the footpoints of the flare loops \citep[also shown by][]{Battaglia2024}. This indicates that thick-target emission is the dominant process, so it remains open why the lower-energy electrons in our sample resemble more closely a thin-target relation which is not expected. 

 In this context, Fig.~\ref{Correlation_delta70-200}b provides an alternative picture, where only in-situ spectral indices evaluated above the first spectral break are considered (i.e., at 110 keV). Each event in this representation is now described by a single point, avoiding a mixing of indices from different spectral regimes. In contrast to Fig.~\ref{Correlation_delta70-200}a, 6 out of our 8 events in Fig.~\ref{Correlation_delta70-200}b lie within the range populated by the sample of \citet{Krucker2007}, between the thin- and thick-target lines.
 Among these 6 events, 4 events lie closer to the thick-target model. One event lies above the thick-target line and only one event remains clearly below the thin-target line in Fig.~\ref{Correlation_delta70-200}b. Notably, this event corresponds to the most intense and hardest in-situ spectrum in our sample at high energies (i.e., >200 keV), suggesting that particular acceleration or transport conditions may have applied in this case.

%% -------------------------------------- %%
%  DISCUSSIONS
%% -------------------------------------- %%

\section{Discussion}
\label{sec:discussion}
%----------

    \subsection{SEE timing and effective path lengths}
    \label{sub:diss-path}
    
In most cases, the VDA results are consistent with the expectations: the inferred injection times broadly coincide with the duration of the nonthermal HXR emission observed by STIX, and the fitted path lengths lie within about 20\% of the nominal Parker spiral values, consistent with moderate scattering. The 3~February~2022 event is a notable exception. Its VDA fit returns a path length almost four times the Parker spiral estimate producing a large discrepancy between the TSA- and VDA-derived injection times (Table \ref{Table:TSA_VDA}). Yet the VDA injection time still matches the HXR peak within uncertainties, suggesting that the extended path may be physical rather than an artefact. Similar cases have previously been reported \citep[e.g.][]{Wimmer2023, Rodriguez-Garcia2021, Rodriguez-Garcia2025a}. Notice that this event is also relatively weak and observed in the EPT–Sun telescope, with a restricted VDA energy range of 31.2–110.7~keV. For this reason, STEP measurements were additionally consulted. Both STEP-only and combined STEP+EPT–Sun fittings were used to perform supplementary VDAs (not shown here). The combined STEP+EPT–Sun fit yields a path length of $3.49 \pm 0.07$~au, while a STEP-only fit gives a shorter value of $2.71 \pm 0.13$~au, still far above the Parker spiral estimate. These independent fits reinforce the interpretation that the unusually long path length is likely physical.

For completeness, we note that the 24~December~2022 event could not be reliably analysed with VDA due to contamination of high-energy electrons in the lower-energy channels, leading to an unphysical path length already reported in \citet{Warmuth2025}. In this case, we relied on the TSA-derived injection time, which remains consistent with the HXR peak. 

   \subsection{In-situ spectral breaks} \label{subsec:is-spectral-breaks}
Broken power-law spectra with well-defined breaks have been widely reported in earlier SEE studies \citep[e.g.][]{Krucker2009,Wang2012,Dresing2020, Rodriguez-Garcia2023a, fedeli2026}. For the five events in our sample fitted with a broken power law with a single spectral transition, the mean break energy of $E_b = 67 \pm 13$~keV is similar within uncertainties to the $\sim$60~keV break reported by \citet{Krucker2009}. While such breaks may carry signatures of the underlying acceleration process, several transport-related mechanisms have also been proposed to account for them, indicating that the observed spectral structure likely reflects a combination of source and propagation effects. 

Previous studies have shown that beam–plasma interactions can naturally produce such spectral signatures. Simulations by \citet{Kontar&Reid2009} and \citet{Reid&Kontar2010,Reid&Kontar2013}, in which an initially single power-law electron population propagates through a non-uniform solar-wind plasma while self-consistently generating and absorbing Langmuir waves, demonstrate that wave–particle interactions alone can produce a clear spectral break: the spectrum flattens below the break and steepens above it, with the break energy forming at a few to several tens of keV, which is consistent with the range found in our sample. In this framework, the steepening above the break does not require a second accelerated population but arises from transport effects imprinted by resonant interactions with the background plasma. Observational support comes from \citet{Lorfing2023}, who analysed energetic electron fluxes, pitch-angle distributions, associated Langmuir waves, and type~III radio bursts. They found that velocity dispersion and quasi-linear relaxation in the deca-keV range, co-temporal with locally generated Langmuir waves, directly shape the observed electron spectrum around $\sim$10–50~keV, in agreement with the simulations previously mentioned.

By contrast, \citet{Dresing2020} reported a mean break energy of $\sim$120~keV, significantly higher than both our mean break at $67 \pm 13$~keV and the $\sim$60~keV found by \citet{Krucker2009}. One possible explanation for this discrepancy can be that some of our events have low statistics, so the high-energy break may have been lost in the background noise. Moreover, breaks at higher energies (E > 100 keV) are likely to be affected by additional transport effects, such as energy-dependent pitch-angle scattering, as suggested by \citet{Strauss2020}, which can modify the high-energy part of the spectrum and shift the apparent break to larger energies, which for our microflares may be below the background level.

The 24~December~2022 event (Fig. \ref{fig:enu-spec-is}c), which required a double broken power-law fit, seems to be a case where several processes influence the spectrum at different energies. Its low-energy break ($23 \pm 2$~keV) falls within the regime where Langmuir-wave interactions are expected to dominate \citep[e.g.][]{Reid&Kontar2013}, while its high-energy break ($284 \pm 68$~keV) is notably large compared to the values found by \citet{fedeli2026}, where the majority of their events were fitted with a double broken power-law. 
By contrast, we find that the 23~December~2022 (Fig. \ref{fig:enu-spec-is}b) event is fitted with a single power law. This agrees well with the findings reported by \citet{Lorfing2023} for low-intensity SEE events. 
However, this is not the case for the 12 November 2022 event (Fig. \ref{all-is-spectra}e). Although its spectrum is also fitted by a single power-law, this event is not associated with weak electron fluxes. The combined STEP+EPT-Sun spectrum exhibits a flattening below $\sim$20~keV, coincident with a drop in pitch-angle coverage around the peak times of these lower STEP energies. While these channels still satisfy the $3\sigma$ enhancement criterion defined in Sect.~\ref{is-data}, this coverage drop likely affects the measured intensities. We therefore restricted the fit to energies above 20~keV. 

It is worth noting that \citet{fedeli2026} analysed a sample of the most 50 intense SEE events in the CoSEE-Cat \citep{Warmuth2025}, typically much stronger than the HMF-associated events considered in our study. This aligns with the approaches adopted by \citet{Krucker2007}, \citet{Krucker2009}, and \citet{Dresing2021}, where the most intense prompt SEE events, with the best data quality and counting statistics, were selected for their statistical studies. This differs from the approach adopted here, where events were selected based on their characteristics at the Sun (i.e., consistent with the definition of HMFs), independently of in-situ observation statistics.  
This methodological difference may therefore have an influence in the comparison between studies.

Due to instrumental effects, artificial spectral breaks can also be introduced, which requires careful consideration. For three events in our sample, 3 February 2022 (Fig. \ref{fig:enu-spec-is}a), 12 November 2022 (Fig. \ref{all-is-spectra}e) and 24 December 2022 (Fig. \ref{fig:enu-spec-is}c), EPT–Sun measurements were combined with STEP observations which extended the spectral energy range. However, such combinations must be treated with caution. As discussed in Section~\ref{is-data}, MAG data were used to determine the magnetic-field direction at the time of the peak flux in each energy channel. When the beam pointed outside the shared STEP–EPT–Sun FOV (i.e could lie within STEP FOV, but outside EPT-Sun FOV), the corresponding channels may not have sampled the main beam and could introduce artificial spectral features, potentially mimicking or shifting spectral breaks. Appendix \ref{app:appendixA3} provides maps of the STEP pixel response with the EPT–Sun FOV overlaid. These maps aim to show where the beam intersects the detectors across the fitted energy range.
Our analysis provides additional context for interpreting the diversity of the combined STEP and EPT–Sun spectra. 

As detailed in Appendix~\ref{app:D1}, for the 24~December~2022 event (panel (c) in Fig.~\ref{fig:enu-spec-is}), the magnetic-field direction at the peak times of all fitted STEP and EPT–Sun channels remains within the shared FOV, consistent with both instruments sampling the main electron beam and supporting a physical origin for the two observed spectral breaks.
For the 3~February~2022 event (panel (a) in Fig.~\ref{fig:enu-spec-is}), by contrast, the magnetic-field directions corresponding to the STEP channels included in the fit are more scattered, with a change in orientation near the $\sim$20~keV range where the spectral break appears. This suggests that the break may be influenced by viewing geometry rather than reflecting a purely physical transition. However, whether this break is of physical or instrumental origin does not impact the results shown in Fig.~\ref{Correlation_delta70-200} for this event, as the spectral indices at 70, 110, and 200~keV remain unchanged for this event. If the spectrum is instead described by a single power law, the mean break energy inferred from the remaining broken power-law fits becomes $79 \pm 14$~keV, which, within uncertainties, remains consistent with the $\sim$60~keV reported by \citet{Krucker2009} and is closer to the mean value of $74.3 \pm 1.6$~keV found for broken power-law spectra in \citet{fedeli2026}.
For the 12~November~2022 event (panel (e) in Fig.~\ref{all-is-spectra}), the magnetic-field direction remains within the shared FOV up to $\sim$28~keV but deviates at higher STEP energies. However, the spectrum associated to this event is well described by a single power law, with no evidence for a spectral break in this energy range. This suggests that, for this event, the observed spectral shape is unlikely to be strongly affected by viewing-geometry.

\subsection{Comparing HXR and in-situ spectral indices}

Our results have shown that the in-situ electron spectra associated with hard microflares in our sample are relatively hard which is consistent with the hard HXR spectra observed for the same events. This reinforces the fact that HMFs efficiently accelerate electrons to high energies \citep{Battaglia2024}, and is now supported by direct in-situ measurements.   
The spectral indices have been compared to earlier studies by \citet{Krucker2007} and \citet{Dresing2021} which include larger samples of flares with a wide range of GOES-classes. 

Similar to these earlier studies, we try to relate the combined HXR and electron spectral indices to the thick-target and thin-target model. In Fig.~\ref{Correlation_delta70-200} we present two complementary analysis schemes for sorting out these spectral indices in relation to the two models.  
In Fig.~\ref{Correlation_delta70-200}a the electron indices are evaluated at 70 and 200~keV for comparability with the study by \citet{Dresing2021} where the analysis was conducted at precisely the same energies. This comes with the feature, that several events can be analysed at both energy regimes. 
While most of spectral indices fall in between the thick-target and the thin-target model, a few points show noticeable deviations at both 70 and 200~keV in Fig.~\ref{Correlation_delta70-200}a. 
Given the small size of our sample, these deviations should not be over-interpreted and they may also partly reflect instrumental differences between the datasets. The Wind/3DP measurements used in \citet{Krucker2007} and the STEREO/SEPT observations used in \citet{Dresing2021} cover a similar energy range as EPT, but at lower energy resolution. This may contribute to the divergence from earlier studies, but it does not facilitate the theoretical explanation of several outliers clearly below  the thin-target model line. In general, it remains unclear why a few of our points in Fig.~\ref{Correlation_delta70-200}a, mainly representative of the lower-energy electrons, appear to lie closer to the thin-target model when the observational evidence of flare footpoints in HXRs points clearly towards the thick-target model.

When extending the analysis to an alternative reference energy at 110~keV (Fig.~\ref{Correlation_delta70-200}b) the situation becomes a bit clearer.
First, we note that each of our HMF-associated events reach up to these energies so that all eight events are represented exactly once in Fig.~\ref{Correlation_delta70-200}b - avoiding any bias in the sample. Second, we ensured by definition that for all events where a spectral break is observed, this energy of 110~keV is above the (first) spectral break. This follows the same theoretical arguments as in \citet{Krucker2007}, where the range above the spectral break is referred to be more likely the representative one for the initial electron acceleration at the flare site.   
As discussed in \ref{subsec:is-spectral-breaks}, we find that six out of the eight events in Fig.~\ref{Correlation_delta70-200}b fall into the regime between the thick-target and the thin-target model, widely consistent with the events reported in \citet{Krucker2007}. Furthermore, in Fig.~\ref{Correlation_delta70-200}b more events are found to lie closer to the thick target model which is in better agreement with the HXR source observations (shown in Fig.~\ref{fig:overview-figure}) as compared to Fig.~\ref{Correlation_delta70-200}a. Interestingly, the only event, that is clearly below the thin target model, is the one from 24 December 2022, which shows the most intense and hardest spectrum at high energies (beyond 200~keV). It might be interesting to investigate whether any special conditions for the acceleration or release of electrons might have occurred at this flaring event, but this is beyond the scope of this current study.  

As discussed previously, the in-situ electron spectra can be altered by energy-dependent transport effects: wave–particle interactions modify the low-energy range, while pitch-angle scattering increasingly affects the highest energies. This implies that the intermediate part of the spectrum is most likely to preserve the characteristics of the injected population at the Sun. This is supported by the findings by \citet{fedeli2026}, who compared in-situ spectral indices with the GOES X-ray peak flux of the associated flares across three reference energies (10, 70, and 200~keV) and found the strongest correlation at 70~keV. Taken together, these earlier results suggest that the spectral indices at intermediate energies (around $\sim$70~keV), provide the most reliable diagnostic for comparing in-situ electrons with their HXR-producing counterparts, as they are least affected by transport distortions and best constrained by current instrumentation. For our sample, a direct comparison between Figs.~\ref{Correlation_delta70-200}a and \ref{Correlation_delta70-200}b suggests that the correlation with HXR spectral indices appears more clearly at 110~keV, with fewer events deviating from the overall trend, than at 70~keV. In this sense, our results show that the clearest correlation is found within an energy range that is (only) slightly higher than the intermediate energy ($\sim$70~keV) identified by \citet{fedeli2026}, but at the same time lies above the spectral break of all measured in-situ electron spectra. 
Despite the small sample size of our study, we believe that this establishes further evidence for determining the relevant energy regime  when relating in-situ electrons to HXRs from flaring sites. Identifying a slightly more energetic in-situ electron  range here compared to \citet{fedeli2026}, might lie in the fact that we are focusing on flares with particular hard spectra - despite being microflares.  

%----------

    \subsection{Hard microflares in previous studies}
    
%%%%BEGIN FIGURE
  \begin{figure}
        \centering
        \includegraphics[width=\linewidth]{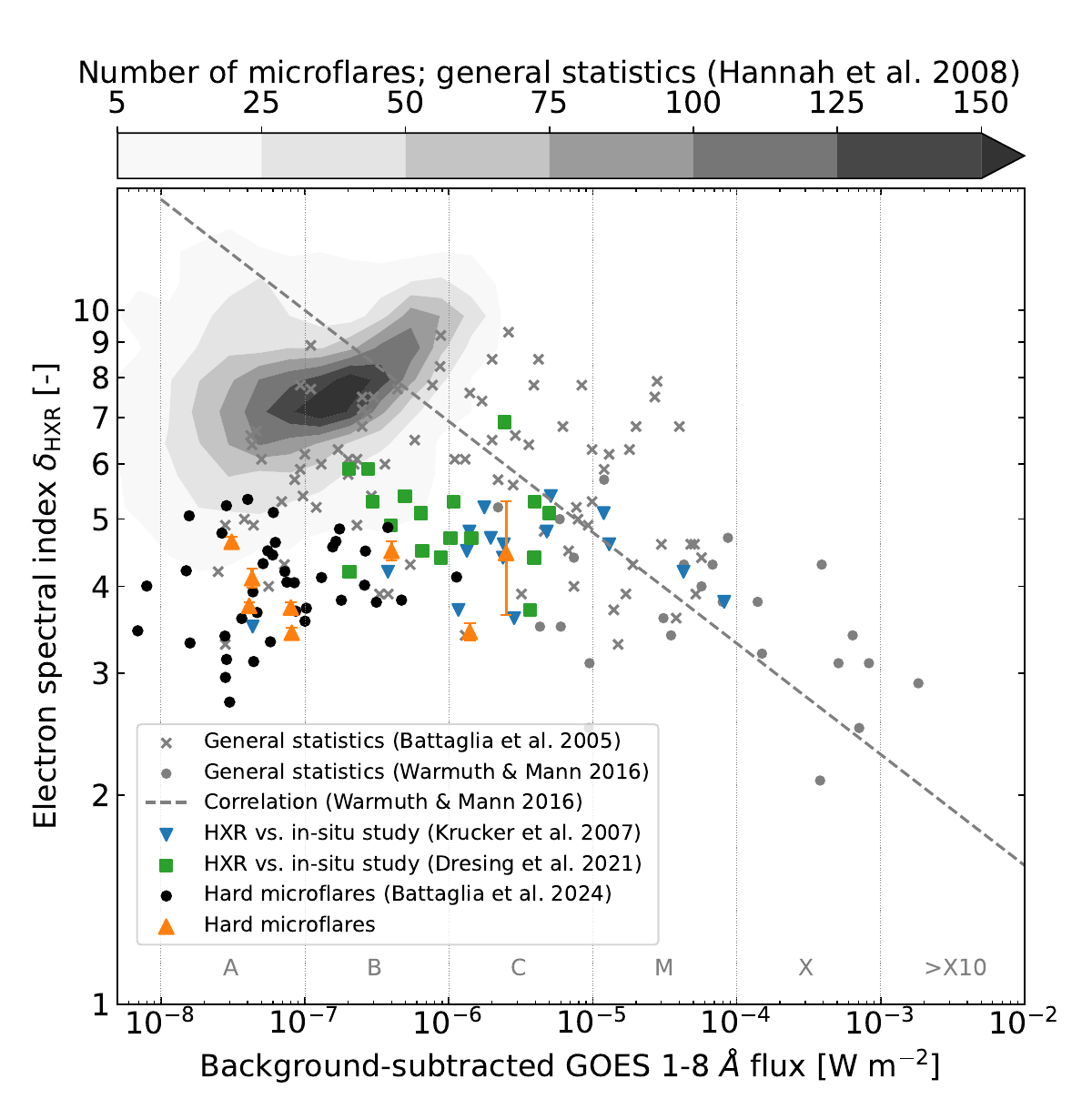}
        \caption{Electron spectral index $\delta_{HXR}$ as a function of the background-subtracted GOES flux of different studies. Events with general statistical studies of flares and microflares have been represented in gray, and the reference sample of hard microflares of \citet{Battaglia2024} in black.  Studies comparing electron spectra in-situ and in HXRs have been represented in blue \citep{Krucker2007} and green \citep{Dresing2021}. The events of this study are included as orange triangles.}
        \label{fig:HXRs-vs-GOES}
    \end{figure}
%%%%END FIGURE  
    As previously discussed, some studies have already reported correlations between flare-accelerated electron distributions derived from HXR observations and in-situ SEE measurements \citep{Krucker2007,Krucker2009,Dresing2021}, although direct comparison remains challenging due to IP transport effects.
    Other studies investigated the source regions at the Sun of in-situ observed electrons to explain possible relationships between the observed electron distributions and the acceleration processes that occurred at the Sun \citep[e.g.][]{Krucker2011,Wang2023}. From this perspective, one of the main mechanisms explaining how these electrons can access IP space is the so-called interchange reconnection model \citep[e.g.][]{Parker1973,Shibata1992,Baker2009}, where emerging magnetic loops reconnect with open field lines.

    As it was done in \citet{Battaglia2024}, in Fig.~\ref{fig:HXRs-vs-GOES} we report the electron spectral index obtained from HXRs, assuming the thick-target model, as a function of the background-subtracted GOES flux at the nonthermal peak time. We can clearly see that our events fall well within the HMF regime defined in \citet{Battaglia2024}, with the events of this study being harder than regular microflares \citep[e.g.][]{Hannah2008}. In addition, we checked previous statistical studies comparing in-situ and HXR spectra \citep{Krucker2007,Dresing2021} in order to investigate which type of events they detected. The first thing to note is the lack of GOES X class flares in the sample of events from \citet{Krucker2007} and \citet{Dresing2021}. The two studies combined\footnote{We note that in both studies of \citet{Krucker2007} and \citet{Dresing2021} no pre-flare signal has been subtracted from the GOES flux. In order to properly compare the events and emphasize only the flare-related signal, we subtracted the pre-flare signal. This is consistent with the approach taken in our work and in \citet{Battaglia2024}.} comprise a total of 33 events: 4 M class flares \citep[all from][]{Krucker2007}, 18 C class, 10 B class, and one A class. Interestingly, the only A class event is also the hardest event in the entire sample (for both HXR and in-situ observations), despite the presence of 4 M class flares. Similarly, the smallest event from \citet{Dresing2021}, which is of B class, is the second hardest in HXRs in the \citet{Dresing2021} sample (and third hardest in-situ). This is noteworthy because it emphasizes that there must be a mechanism in these microflares that makes the electron acceleration very efficient, a finding confirmed by both HXR and in-situ observations.

    \begin{figure*}
        \centering
        \includegraphics[width=\linewidth]{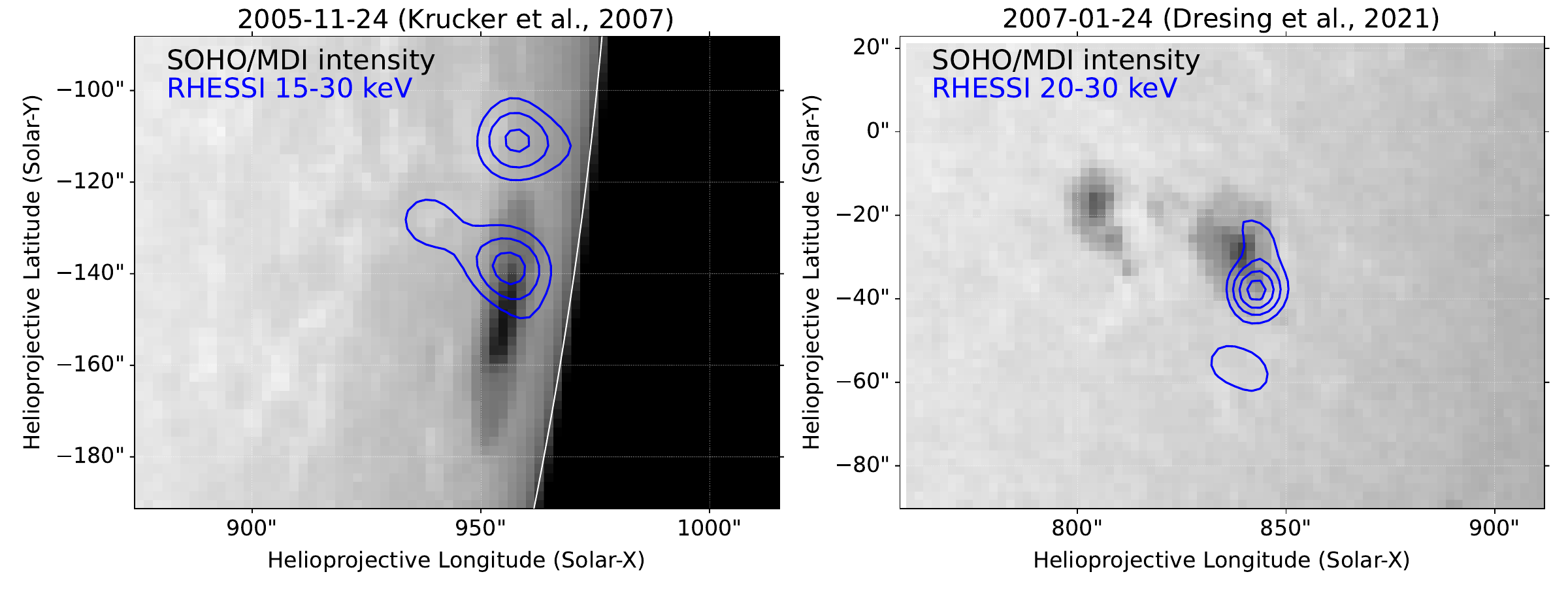}
        \caption{HXR RHESSI images of two events from previous studies. \emph{Left panel}: SOHO/MDI intensity image with RHESSI contours (blue) overlaid for the smallest event reported in \cite{Krucker2007}. The event is clearly rooted in a sunspot. \emph{Right panel}: Similar display for the smallest event in \citet{Dresing2021}. This event is also rooted in a sunspot.}
        \label{fig:images-previous-studies}
    \end{figure*}
    Following the fact that the two smallest events in the combined sample of \citet{Krucker2007} and \citet{Dresing2021} are among the hardest events, we analysed their morphology at the Sun. We examined the HXR observations from the Reuven Ramaty High-Energy Solar Spectroscopic Imager \citep[RHESSI;][]{Lin2002} and compared them with visible light observations from the Michelson Doppler Imager \citep[MDI;][]{scherrer1995mdi} aboard the Solar and Heliospheric Observatory \citep[SOHO;][]{Domingo1995}, following the approach of \citet{Battaglia2024} and \citet{Saqri2024}. Figure~\ref{fig:images-previous-studies} shows this comparison. This figure clearly shows that one footpoint is rooted in a sunspot, which is consistent with our findings and previous studies of HMFs \citep{Battaglia2024,Saqri2024}. These events are among the smallest yet simultaneously the hardest, and both share the common feature of being rooted in sunspots. This highly efficient electron acceleration process is therefore likely related to either the sunspot magnetic field strength or its morphology. Since open magnetic field lines are required to observe electrons in-situ, this type of morphology is a good candidate for being the source of systematic in-situ observations. It is now interesting to investigate whether any relationship exists between hard microflares and the interchange-reconnection model typically invoked to explain in-situ observations of impulsive SEPs. Beyond electron acceleration, it is also intriguing to investigate whether such a mechanism leaves any imprints in in-situ observations of ions \citep[e.g.][]{Mason2025}.

%% -------------------------------------- %%
%  CONCLUSIONS
%% -------------------------------------- %%

\section{Conclusions and outlook}
\label{sec:Conclusions}

Thanks to the spectral and temporal capabilities provided by the STIX and EPD instruments aboard Solar Orbiter, in this work a sample of HMFs is systematically linked to in-situ SEEs for the first time. We compared their spectral shapes and obtained key results that we summarise in the following.

\begin{itemize}
    \item The electron injection times are closely linked with HXRs, with a mean delay of about 4.9 minutes relative to the HXR peak, consistent with values reported for impulsive and prompt SEE events \citep[e.g.,][]{Krucker2007,Warmuth2025}. Corresponding VDA path lengths lie within $\sim 20\%$ of nominal Parker spiral lengths for most events, indicating moderate IP scattering. 
    \item One event (3 February 2022) exhibits an anomalously long path length, nearly four times the Parker estimate, while remaining consistent with the HXR peak within uncertainties, in line with similar cases reported previously \citep[e.g.,][]{Rodriguez-Garcia2021, Rodriguez-Garcia2025a, Wimmer2023}. Supplementary STEP measurements support the physical nature of this long path length. 
    \item STIX observations show HXR photon indices ranging from $\gamma = 2.43 \pm 0.06$ to $\gamma = 3.64 \pm 0.14$, indicating hard nonthermal spectra concistent with \citet{Battaglia2024}. In-situ electron spectra are likewise hard, with mean spectral indices of $2.36 \pm 0.29$ at 70~keV and $3.18 \pm 0.25$ at 200~keV. These HMFs populate the hard end of previously reported flare–SEE distributions \citep{Krucker2007,Dresing2021}, even compared to samples that include larger C- and M-class flares, demonstrating that efficient acceleration to high energies does not necessarily require large flare energy release.
    \item Our HMFs have a footpoint rooted in or at the edge of a sunspot, consistent with \citet{Battaglia2024} and \citet{Saqri2024}. In addition, by revisiting earlier joint HXR–in situ studies \citep{Krucker2007,Dresing2021}, we found that among two of the smallest, yet among the hardest, events in their sample have also a footpoint rooted in a sunspot, suggesting that some HMFs populated their sample. Therefore, the magnetic configuration of HMFs and their proximity to the open magnetic filed lines of the sunspot, may be particularly favourable for enabling high-energy electrons to escape into interplanetary space.
    \item In-situ electron spectra of our HMF events exhibit a range of spectral shapes, including power-laws, broken power-laws, and double broken power-laws. While the broad energy coverage of EPD allows these features to be resolved \citep[e.g.][]{Lorfing2023,fedeli2026}, the physical origin of the observed spectral diversity remains unclear.
\end{itemize}

The history of in-situ observations of impulsive SEEs demonstrates that notable in-situ signatures do not necessarily originate from large flares. Instead, numerous studies show that prominent in-situ detections can be associated with microflare-type of events (e.g. \citet{lario2024novemberevents}, \textcolor{blue}{Janitzek et al. 2026, submitted}).
In this study, we provide a new possible explanation for this counter-intuitive behaviour: hard microflares. Through their efficient electron acceleration and peculiar magnetic morphology, hard microflares may create the right conditions that allow high-energy electrons to escape particularly effectively into the heliosphere. To address this question more rigorously, future work will require extending this type of combined HXR–in-situ analysis to a substantially larger event sample.

A broader perspective involves examining how high-energy electrons accelerated in hard microflares propagate through the heliosphere. Searching for associated type~III radio bursts or Langmuir-wave activity may clarify the link between flare acceleration and the electrons detected in-situ. Additionally, it is also interesting to investigate whether such an efficient acceleration mechanism associated with HMFs might also be a source for suprathermal ions \citep[e.g.][]{Mason2025}.

\begin{acknowledgements} 

Solar Orbiter is a space mission of international collaboration between ESA and NASA, operated by ESA. The STIX instrument is an international collaboration between Switzerland, Poland, France, Czech Republic, Germany, Austria, Ireland, and Italy. IRSOL is supported by the Swiss Confederation (SERI), Canton Ticino, the city of Locarno and the local municipalities.

A.F.B. is supported by the Swiss National Science Foundation (SNSF) grant 200020\_213147. L.R.-G.\ acknowledges support through the European Space Agency (ESA) research fellowship programme.
 R.G.-H. and F.E.L acknowledge financial support under Project PID2023-150952OB-I00 funded by MICIU/AEI/10.13039/501100011033 and by FEDER, EU.
L.R.-G and R.G.-H  acknowledge financial support by EU/FEDER and JCCM/INNOCAM, under project
SBPLY/24/180225/000108.

This study makes use of analysis tools developed within the European Union’s Horizon 2020 research and innovation programme (SERPENTINE).
\end{acknowledgements}

\begin{flushleft}

\textbf{ORCID iDs} 
\vspace{2mm}

Andrea Francesco Battaglia \orcid{https://orcid.org/0000-0003-4490-7344} \\
Laura Rodríguez-García \orcid{https://orcid.org/0000-0003-2361-5510}\\
Francisco Espinosa Lara \orcid{https://orcid.org/0000-0001-9039-8822}

\end{flushleft}

\bibliographystyle{bibtex/aa}
\bibliography{bibtex/biblio.bib}
%\clearpage
%\onecolumn
%\onecolumn

% -------------------------------------- %%
% APPENDIX
% -------------------------------------- %%

\begin{appendix}
\nolinenumbers

\section{VDA analysis: SERPENTINE tool and EPT-viewing selection}
\label{app:appendixA1}
\subsection{VDA method}
Under the standard VDA assumptions described in Section \ref{is-data}, electrons of different energies are released simultaneously at the Sun and propagate scatter-free along a single effective path length $L_{eff}$, with constant speed, given by their energy $E=\gamma mc^2$. The onset times $t_{onset}\left(v_{i}\right)$ at the spacecraft of the energetic electrons follow a velocity dispersion pattern, and are given for each energy channel $i$ by \citep[c.f][]{Vainio2013}: 
\begin{equation}
 t_{onset}\left(v_{i}\right)= t_{srt} + \frac{L_{eff}}{c\beta\left(v_{i}\right)}
 \end{equation}
where $t_{srt}$ is the injection time at the Sun, $L_{eff}$ is the effective path length and $\beta\left(v_{i}\right)=v_{i}/c$, where $v_{i}$ represents the velocity of the electrons associated with energy channel $i$. Notice that $v_{i}$ is computed directly from the kinetic energy $E_i=\sqrt{E_{low,i}E_{up,i}}$, that is taken to be the geometric mean of the corresponding energy channel boundaries for each energy channel $i$ introduced. 

When the onset times at each energy channel, are plotted as a function of $1/v$ (or $c/\beta$) a linear fit yields both the effective path length $L_{eff}$ as the slope of the curve and the injection time at the Sun $t_{srt}$ as the intercept with the y-axis. 
\subsection{SERPENTINE tools: onset time and peak flux determination}
     \begin{figure}
        \centering
         \includegraphics[width=0.48\textwidth]{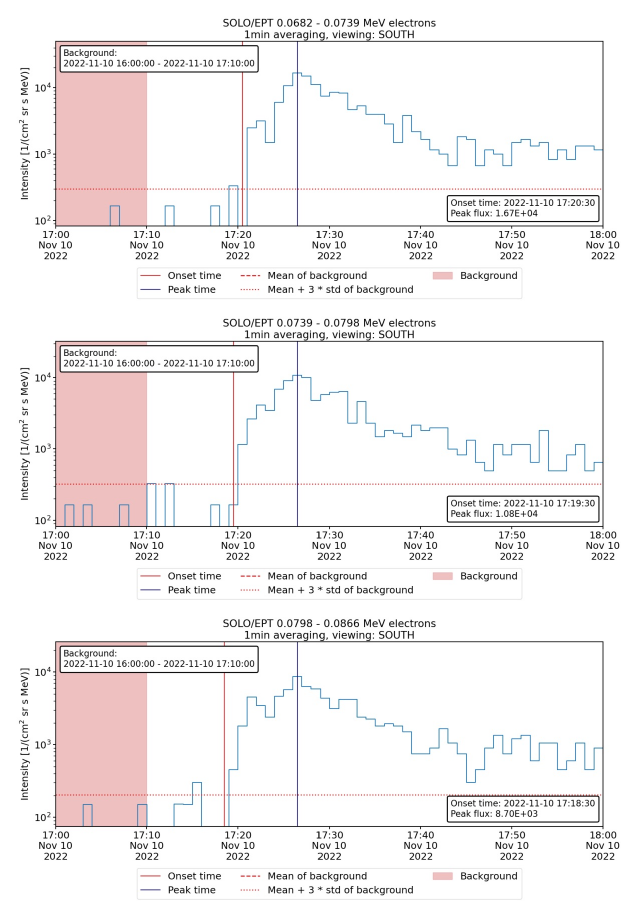}
         \caption{Example of intensity-time series for the event on 10 November 2022, highlighting the determination of both the onset times (vertical red line) and the peak intensities (vertical blue line) for three consecutive energy channels of EPT-South. The light red area is the pre-event background window, used to map the CUSUM parameters (mean and standard deviation of the pre-event background, provided in the legend). The energy range associated to the energy bin, and the averaging used for this event are provided in the title of each plot. }
        \label{fig:timeseries_ex}
      \end{figure}
\cite{Palmroos2025} introduces a suite of tools developed within the Solar EneRgetic ParticlE aNalysis plaTform for the INner hEliosphere (SERPENTINE), that enable automated downloading, processing, and analysis of charged-particle measurements from missions such as Solar Orbiter. For each energy channel, the intensity time series of electrons is first resampled into bins of fixed duration $\Delta$ (typically 1 min), over a selected plotting range and used consistently for both onset detection and peak-flux determination.

The onset-detection routine applies a Poisson-CUSUM procedure, complemented by bootstrap resampling of the pre-event background to estimate uncertainties. The method identifies the earliest statistically significant rise above background levels: a quiet interval is used to compute the background mean and standard deviation, which the CUSUM algorithm updates dynamically as it progresses through the data. An onset is registered once the accumulated CUSUM signal exceeds a threshold corresponding to a flux enhancement above the background. In our analysis we adopted a threshold of mean + 3$\sigma$ (instead of the default 2$\sigma$, see \cite{Palmroos2025}) to ensure a more conservative and reliable onset identification, especially for events with strong pre-event fluctuations. A visual example of the onset determination is shown in Fig.~\ref{fig:timeseries_ex}, where intensity–time profiles for three consecutive EPT–South energy channels of the 10~November~2022 event are displayed. In each panel, the light-red shaded region marks the pre-event background interval used to derive the CUSUM parameters, while the vertical red line indicates the automatically determined onset. The corresponding energies per channel and the chosen time averaging (or resampling) are indicated in each panel.

For the determination of peak fluxes, the SERPENTINE routine operates on the same resampled time series used for the onset analysis. Within a user-defined plotting interval (e.g. 17:00–18:00 in Fig.~\ref{fig:timeseries_ex}), the tool computes, for each energy channel, the maximum flux value in every resampled bin. The peak flux for that channel is then defined as the highest of these flux maxima per resampled bin, over the selected interval. The vertical blue line in Fig.~\ref{fig:timeseries_ex} marks the peak intensity for three different energy channels. We notice that the tool does not yet propagate the corresponding uncertainties associated with these peak flux values. We therefore needed to compute manually the peak-flux uncertainty for each energy channel. Since the peak flux is derived from a resampled bin, its uncertainty was estimated as the standard error of the mean (SEM) of that bin, calculated from the per-sample uncertainties, available in the original time series. This ensures that the uncertainty associated with the resampled peak flux reflects the statistical variability within the bin from which the peak value is taken. 

\subsection{EPT-FOVs selection}
Before performing both VDA and in-situ energy spectrum fitting, it is necessary to determine the optimal viewing direction of the EPT sensor, selecting among the Sun, Anti-Sun, North, and South telescopes.

\begin{figure}[ht!]
    \centering
        \includegraphics[width=\linewidth]{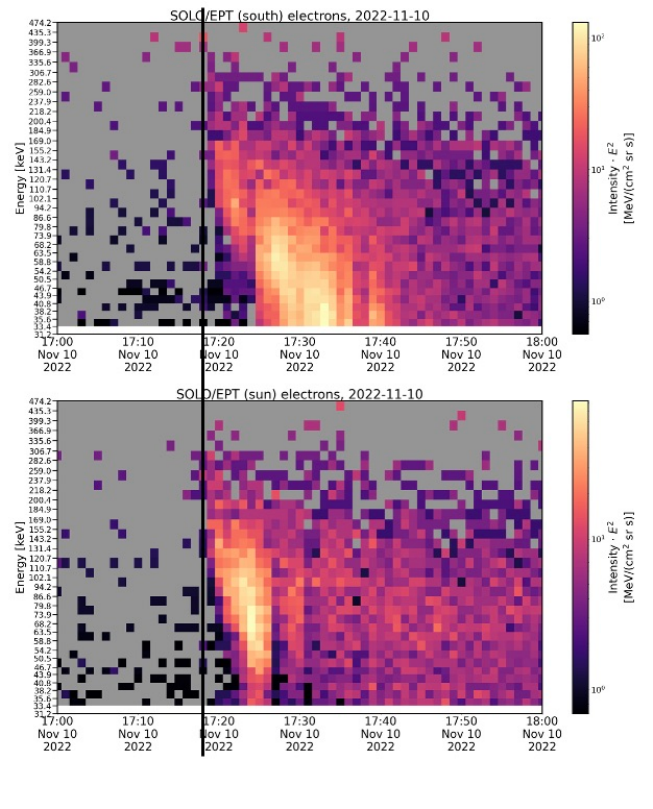}
    \caption{Example of EPD level 2 spectrograms for the event on 10 November 2022, from two EPT viewing directions: South (top) and Sun (bottom). The x-axis shows the time (in UT), the y-axis shows electron energy (in keV), and the color scale indicates electron intensity. The bright diagonal structures trace the electron signal across energy channels, revealing velocity dispersion.}
     \label{fig:Sun+South_221110}
\end{figure}

For VDA fitting, we chose the telescope that detected the first arriving electrons, as this direction is most likely to satisfy the underlying assumptions of VDA (good magnetic connectivity from the start of the event or limited interplanetary scattering). To identify this direction, we inspect dynamic spectrograms such as those shown in Fig.~\ref{fig:Sun+South_221110}, which display electron intensity as a function of time and energy for both South (top panel) and Sun (bottom panel) EPT field of views. The example in Fig.~\ref{fig:Sun+South_221110} shows the event on 10~November~2022, with the SEE onset around 17:15~UT marked by a vertical black line.

For this specific event, however, it is not immediately evident whether the Sunward or Southward telescope detects the earliest electrons: both show a comparable high-energy onset near 17:15~UT, as it is shown in their zoomed spectra in Fig.~\ref{fig:Sun+South_221110}. As the earliest onset cannot be clearly distinguished in this event, we relied on additional diagnostic and ultimately adopted the Southward telescope as the one observing the first arrival particles for two reasons. First, the Sunward telescope exhibits a pronounced drop in peak flux around 17:26~UT at energies near 60~keV in Fig. \ref{fig:Sun+South_221110}, likely caused by a change in pitch-angle coverage, whereas the Southward telescope shows a more stable rise and peak across these energies. Second, the Southward direction displays the highest peak intensity among all EPT FOVs. This is consistent with the selection criterion used for constructing in-situ peak-flux spectra, as described in Sect.~\ref{is-data}. Therefore, for the 10 November 2022 SEE event, we selected the EPT-South FOV for both VDA and spectral analysis, as it shows the most reliable and statistically robust signal.

  Table~\ref{tab:EPT_FOV_selection} lists, for each event, the EPD sensors (STEP, EPT FOVs) selected for the VDA and in-situ spectral fitting. When multiple sensors are listed, they were combined in the corresponding analysis.

 \begin{table}[h]
\centering
\small
\scriptsize
\caption{EPD sensors selected for VDA and TSA, as well as for the in-situ peak-flux spectral fitting for each analyzed event.}
\label{tab:EPT_FOV_selection}
\begin{tabular}{lcc}
\hline
HMF date  & FOVs selected (VDA-TSA) & FOVs selected (spectrum) \\
\hline
9 May 2021        &  EPT-North        & EPT-South   \\
23 May 2021       &  EPT-Sun          & EPT-Sun      \\
3 Feb 2022        & EPT-Sun           & STEP, EPT-Sun \\
10 Nov 2022       & EPT-South         & EPT-South     \\
12 Nov 2022       & EPT-Sun           & STEP, EPT-Sun  \\
23 Dec 2022       & EPT-North         & EPT-North       \\
24 Dec 2022       & EPT-Sun           & STEP, EPT-Sun   \\
15 Nov 2024       & EPT-SUN           & EPT-Sun        \\
\hline
\end{tabular}
\end{table}

\section{TSA and VDA Results}
\label{app:appendixA2}

Table \ref{Table:TSA_VDA} summarises the results of the TSA and VDA analyses performed for the eight HMF–SEE events. TSA parameters (release time, nominal Parker spiral length, and selected energy channel) are listed in Cols. 4–6, while the VDA energy ranges and fitted injection times and path lengths appear in Cols. 7–9.

For TSA, which relies on a single energy channel, we selected a channel with a clear, well-defined intensity enhancement, without significant noise or pre-event fluctuations. For the VDA, as outlined in Sect. \ref{is-data}, we selected all energy channels exhibiting a clear onset, defined as a flux increase exceeding the mean pre-event background by more than 3$\sigma$, ensuring reliable onset determination and a consistent velocity-dispersion pattern.

The nominal Parker spiral length $s_{nom}$ (Col. 4 in Table \ref{Table:TSA_VDA}) represents the shortest travel distance for scatter-free electron propagation and is required for both TSA (to derive the release time for a given energy) and for evaluating the VDA-inferred effective path length. For each event, we computed $s_{nom}$ using the local solar-wind speed measured by Solar Orbiter/SWA and the information of coordinates and separation angles, provided by the Solar MAgnetic Connection HAUS tool \citep[Solar-MACH;][]{Gieseler2023}, which derives and visualizes the spatial configuration and solar magnetic connection of different observers (including Solar Orbiter). The value for the Parker spiral length $s_{nom}$ under these assumptions, is:
\begin{equation}
    s_{nom}=\frac{b}{2}\left[ \theta\sqrt{1+\theta^{2}} + \ln\!\left( \theta + \sqrt{1+\theta^{2}} \right) \right],
\end{equation}
where $b=\frac{r}{\theta}$, with $r$ the heliocentric distance of the spacecraft and $\theta$ the difference between the magnetic footpoint Carrington longitude and the flare Carrington longitude (in radians). 
% \begin{table*}[t]
\begin{table*}[ht!]
  \centering
  \small
  %\begin{ThreePartTable}
   \setlength{\tabcolsep}{8pt}
    % \scriptsize
    \caption{Table summarizing the results from TSA and VDA for the HMF-associated SEE events analysed in this study.}
    \label{Table:TSA_VDA}

    \begin{tabularx}{\textwidth}{@{} cc cccc ccc @{}}
      %\toprule
      \cline{1-9}
      \multicolumn{2}{c}{Hard microflare event} & \multicolumn{4}{c}{SEE event-TSA} &\multicolumn{3}{c}{SEE event-VDA}\\
      \cmidrule(lr){1-2}\cmidrule(lr){3-6}\cmidrule(lr){7-9}
      Date & HXR peak time$^{\mathrm{start}}_{\mathrm{end}}$&
      $R$ & $s_{nom}$  & Energy range & $t_{\mathrm{srt-TSA}}$ & Energy range &$t_{\mathrm{srt-VDA}}$&$L_{\mathrm{eff}}$\\
      %\hline
      \cline{1-9}
          & (UT) & (au) & (au) &(keV) & (UT)&(keV) &(UT)& (au)  \\ (1)&(2)&(3)&(4)&(5)&(6)&(7)&(8)&(9)\\
      %\midrule
      \cline{1-9}

    2021/05/09 & 13:45:59$^{13:41:23}_{13:53:23}$ & 0.92 & 1.10 & 54.2 - 58.8 & 13:52:41$\pm$0:01:00& 35.6 - 282.6 &13:48:30$\pm$0:01:25 & $1.35\pm0.08$ \\
    
    2021/05/23 & 04:26:32$^{04:26:00}_{04:30:20}$ & 0.95 & 1.17 & 54.2 - 58.8 & 04:33:50$\pm$0:01:00 & 35.6 - 282.6 &04:32:47$\pm$0:01:25 & $1.15\pm0.08$\\
    
    2022/02/03 & 20:33:52$^{20:31:20}_{20:38:04}$ &0.83 & 0.89 & 35.6 - 39.6 & 21:31:49$\pm$0:01:00 &31.2 - 102.1 &20:38:09$\pm$0:03:59 & $3.22\pm0.21$ \\
    
    2022/11/10 & 17:06:28$^{17:05:08}_{17:10:20}$ & 0.61 & 0.65 & 46.7 - 50.05 &  17:09:42$\pm$0:01:00&31.2 - 200.4 &17:08:43$\pm$0:01:24 & $0.66\pm0.08$ \\
    
    2022/11/12 & 04:24:42$^{04:24:14}_{04:26:49}$ & 0.63 & 0.69 & 54.2 - 58.8 & 04:28:32$\pm$0:01:00 &31.2 - 200.4 &04:27:42$\pm$0:01:24 & $0.73\pm0.08$\\
    
    2022/12/23 & 04:26:28$^{04:25:36}_{04:32:00}$ & 0.92 & 0.99 & 46.7 - 50.05 & 04:36:45$\pm$0:01:00&31.2 - 102.1 &04:36:14$\pm$0:01:58 & $1.07\pm0.10$ \\
    
    2022/12/24 & 04:02:51$^{04:00:51}_{04:08:11}$ & 0.93 & 1.09 & 237.9 - 259.0 & 04:00:14$\pm$0:01:00 &N/A &N/A & N/A\\
    
    2024/11/15 & 20:21:57$^{20:21:37}_{20:22:29}$ & 0.78 & 0.89 & 54.2 - 58.8 &20:29:27$\pm$0:01:00&31.2 - 184.9 &20:28:22$\pm$0:01:06 & $0.91\pm0.06$ \\
       \cline{1-9}
      %\bottomrule
        \end{tabularx}
    
      %\end{ThreePartTable}
     
      \footnotesize{\textbf{Notes.} Col. 1--2: STIX HXR peak time at the Sun, corrected for the light travel time of the photons. Upper and lower indices in Col. 2 represents the start and end time of the nonthermal HXR emission, respectively. Col. 3: Radial distance from the Sun to Solar Orbiter. Col. 4: Nominal Parker spiral length. Col. 5: Energy range of the electrons used for TSA. Col. 6: TSA injection time along with its error. Col. 7: Energy range of the electrons used for VDA fitting. Col. 8-9: VDA release time and effective path length, respectively.}

% \end{table*}
\end{table*}

\section{Additional HXR and In-situ spectra}
\label{additional-spec}
\begin{figure*}[ht!]
        \centering
         \includegraphics[width=\linewidth]{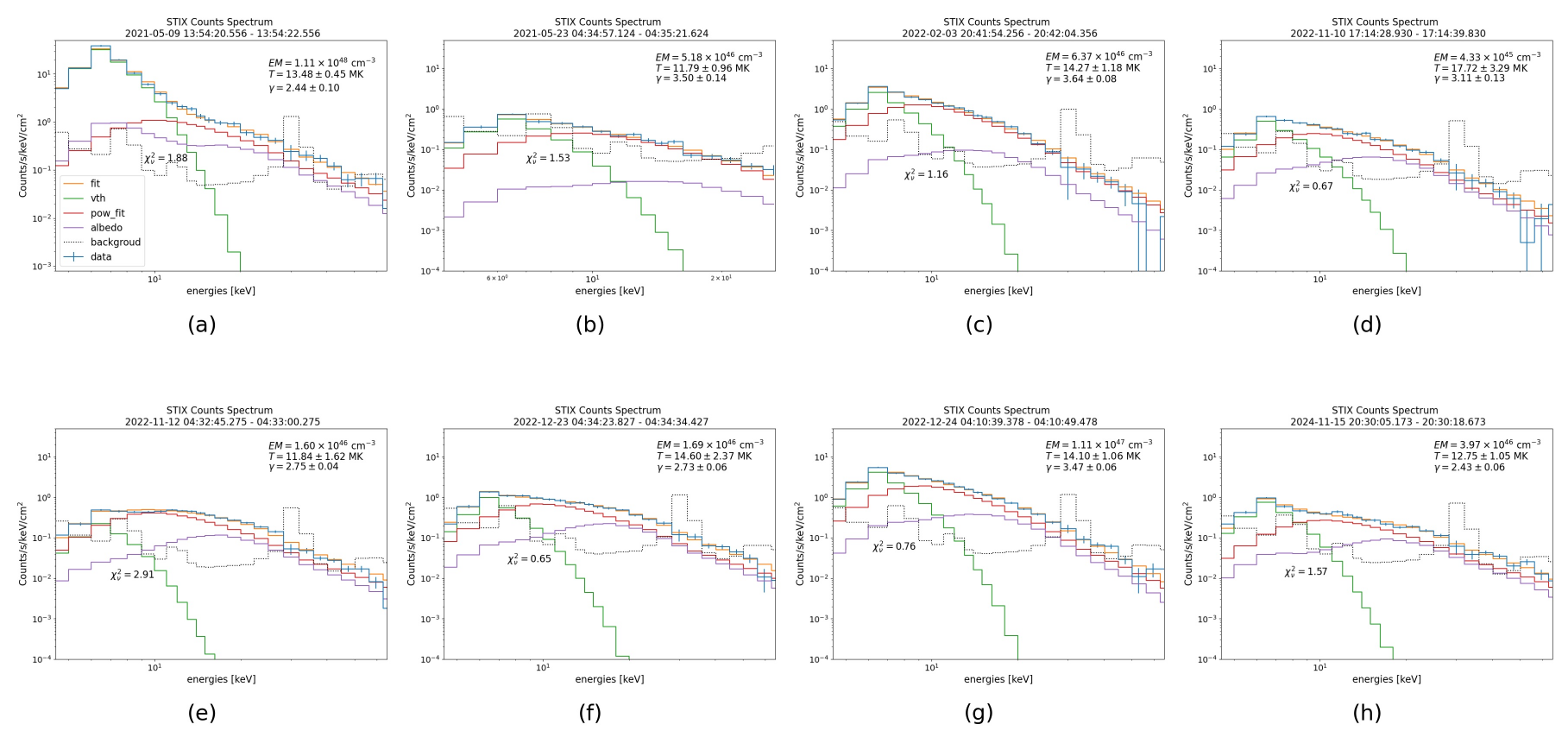}
         \caption{Pre-flare background subtracted STIX count spectra (blue) for all HMF events analysed in this study (panels a to h). The spectra at low energies are fitted with an isothermal fit (green curve), and at high-energies, with a single power-law (red curve), $\gamma$ is the photon spectral index. The albedo component (purple curve) is added for each fitting to account for HXRs reflected from the solar surface. For each event, the total spectrum is shown in orange and the subtracted background is shown in dotted black.}
        \label{all-HXR-spectra}
      \end{figure*}

\begin{figure*}[ht!]
        \centering
         \includegraphics[width=\linewidth]{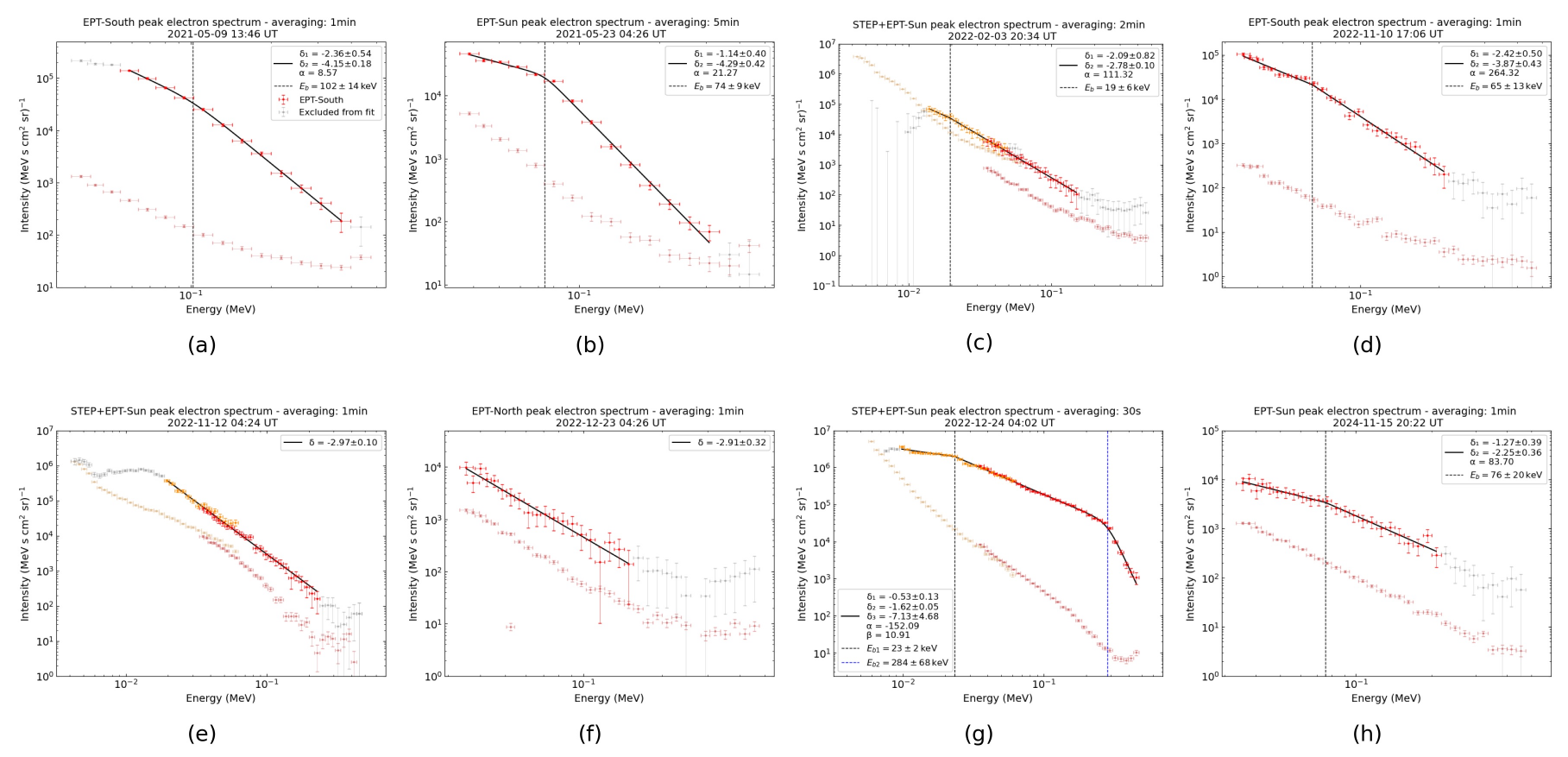}
         \caption{Background-subtracted EPD peak flux spectra for all events analysed in this study (panels (a) to (h)). The fainter and lower intensity points show the pre-event background, that is subtracted from the peak intensities. The gray points denote energy channels that were excluded from the fit. The orange and red points represent STEP and EPT-Sun electrons, respectively. The dashed vertical black lines represent the spectral break of the broken power-law fittings and the first break of the double broken power-law fitting (panel (g)). The blue line in panel (g) represent the second spectral break of the double broken power-law fitting. The legend in each panel provides the fit parameters.}
        \label{all-is-spectra}
      \end{figure*}

    This Appendix presents the STIX HXR count spectra in Figure \ref{all-HXR-spectra} and the EPD in-situ electron peak-flux spectra in Figure \ref{all-is-spectra} for all eight hard microflare events analysed in this study. Both figures show the full spectral fits for each event, including the thermal and nonthermal HXR components (Fig. \ref{all-HXR-spectra}) and the background-subtracted in-situ electron spectra with the corresponding fit models and excluded channels (Fig. \ref{all-is-spectra}).
\section{Combining STEP ad EPT-Sun FOVs}
\label{app:appendixA3}
\subsection{Geometric differences between STEP and EPT sensors}

\begin{figure}
        \centering
         \includegraphics[width=0.48\textwidth]{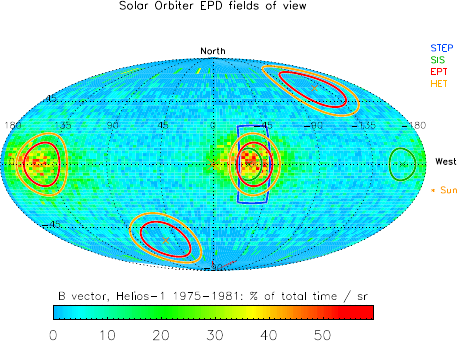}
         \caption{Fields of view of the EPD sensors in the spacecraft reference
frame from \citet{Rodriguez-Pacheco2020}, where the colour-coded background corresponds to the hourly averaged magnetic field vector distribution as observed by HELIOS-1.}
        \label{fig:EPD-FOVs}
    % \end{figure*}
      \end{figure}
      
      \begin{figure}
        \centering
         \includegraphics[width=0.40\textwidth]{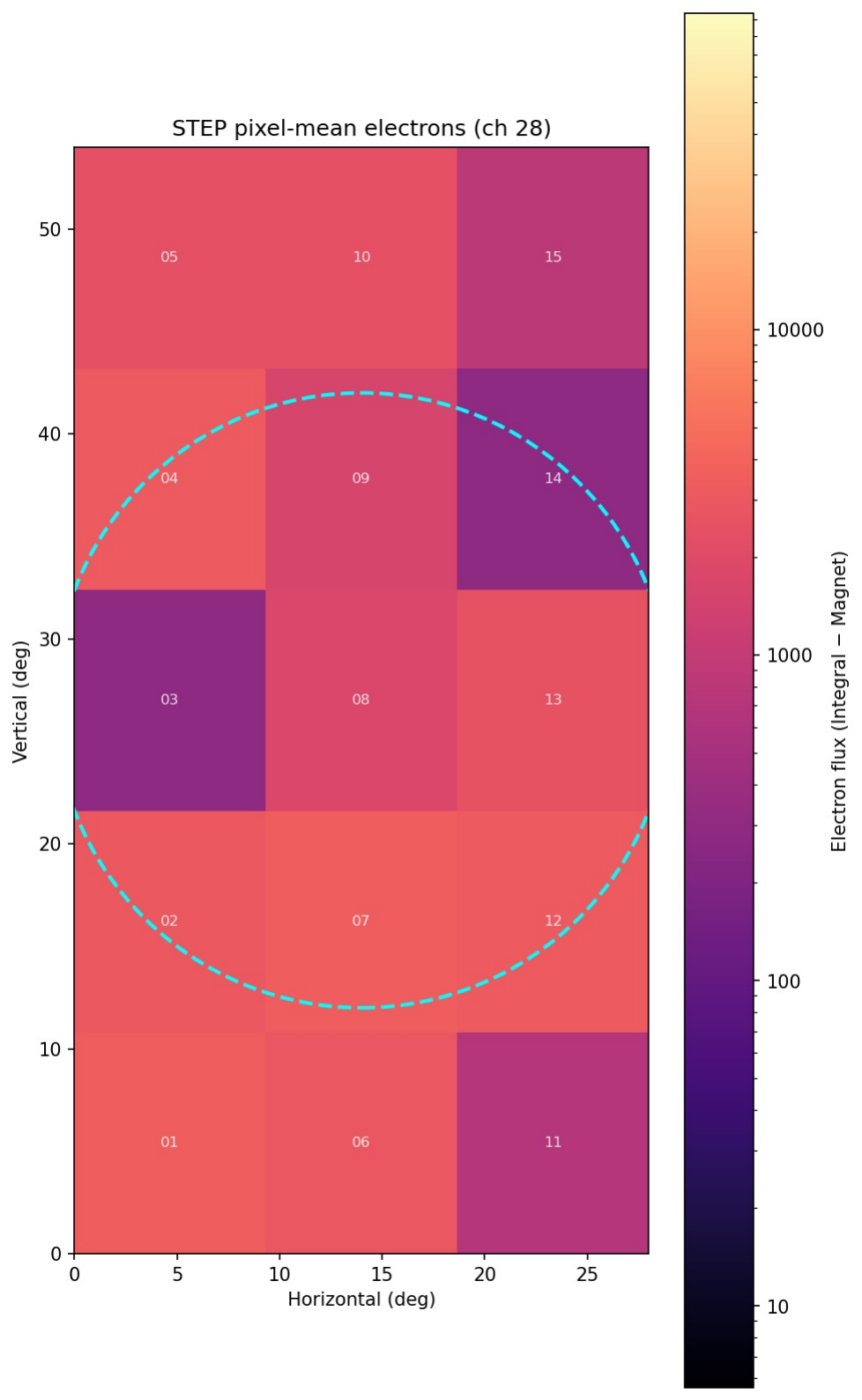}
         \caption{Schematic of the FOV of STEP sensor for the STEP energy channel 28 of the 3 February 2022 event. This map represents the STEP rectangular instrument opening of $28^{\circ}\times54^{\circ}$, divided into its 15 main pixels. The blue circle represents the EPT-Sun FOV. The vertical colour bar accounts for the time-averaged STEP electron flux, in each pixel and on a logarithmic scale.}
        \label{fig:step_ept_map}
    % \end{figure*}
      \end{figure}
As mentioned in Sect. \ref{is-data}, whenever an event is observed by the Sunward telescope of EPT, complementary measurements from STEP can be considered because both instruments nominally observe in the same direction. This can be seen in Fig. \ref{fig:EPD-FOVs} taken from \citet{Rodriguez-Pacheco2020}. In our standard analysis, STEP fluxes are extracted from the \texttt{Integral\_Avg\_Flux} product, which averages the signal over all STEP pixels. However, for highly anisotropic SEP events, the larger FOV of STEP ($28^\circ$ × $54^\circ$ per aperture) compared to the narrower EPT–Sun FOV ($30^\circ$ diameter, shown in Fig. \ref{fig:EPD-FOVs}) can lead to significant discrepancies in the measured intensities for the overlapping energy channels. In such cases, STEP may sample electron beams in angular regions not covered by EPT-Sun, producing intensity offsets and potentially influencing the inferred spectral shape.

To account for these geometric differences, and following an approach similar to that of \citet{fedeli2026}, we used maps such as Fig.~\ref{fig:step_ept_map}. The figure illustrates the STEP FOV for a given energy channel as an example (here channel~28 of the 3~February~2022 event), where the colour scale accounts for the time-averaged STEP electron flux in each pixel on a logarithmic scale. For simplicity, the STEP detector is represented by its 15 main pixels, which together span the full rectangular aperture of $28^\circ \times 54^\circ$. Notice that the smaller edge pixels in between are neglected in this representative method. The EPT-Sun FOV is overlaid as a dashed blue circle as seen in Fig.~\ref{fig:step_ept_map}, allowing the identification of STEP pixels, or fractional pixel areas, that fall within the EPT-Sun FOV. These geometric fractions are then used as weights to compute a weighted spatial average of the STEP flux, reducing the intensity discrepancies in the overlapping energy range, particularly for highly anisotropic events, where the beam is best seen and detected by the outer STEP pixels. The weighted spatial average of the STEP flux is therefore used in this study, to obtain combined STEP-EPT spectra, as shown in Fig. \ref{fig:enu-spec-is}, panels (a) and (c) and in Fig. \ref{all-is-spectra}, panels (c), (e) and (g).

\subsection{Assessment of Interplanetary Magnetic Field (IMF) orientation with respect to the EPD sensors}
\label{app:D1}
SEEs enter the EPD sensors primarily along the IMF. To assess whether the in-situ electron spectra sample the main particle beam, we examined magnetic-field measurements from MAG in conjunction with STEP and EPT–Sun observations. As discussed in Sect.~\ref{is-data}, the in-situ magnetic-field direction at the time of the peak electron flux in each energy channel provides a key diagnostic of whether the detected electrons originate from within the FOV of both particle sensors.

\begin{figure*}[ht!]
    \centering
        \includegraphics[width=\linewidth]{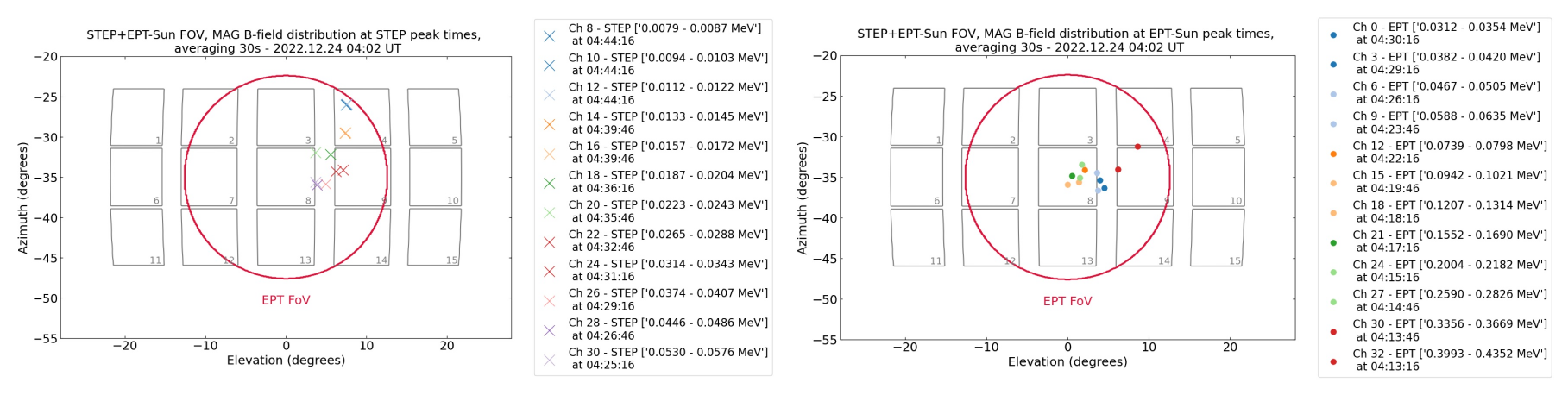}
    \caption{Sky-projected representation of the FOV of STEP sensor, with its pixels mapped into elevation-azimuth coordinates, for the 24 December 2022 SEE event. The EPT-Sun FOV is overlaid with a red circle in each panel. The figure outlines the magnetic field measurements averaged over time with the same resampling (i.e 30 seconds) as applied to the STEP (left panel) and EPT-Sun (right panel) electron flux when determining peak intensities.}
    \label{fig:MAG-STEP+EPT_FOVs_20221224}
\end{figure*}

For this purpose, we used a sky-projected representation of STEP FOV, in which the pixels are mapped into elevation–azimuth coordinates, along with the EPT–Sun FOV (shown in Figs.~\ref{fig:MAG-STEP+EPT_FOVs_20221224} to ~\ref{fig:MAG-STEP+EPT_FOVs_20221112}). In this angular frame, we note that the Sunward direction ($+X$) lies at the top and the North ($+Z$) to the right, in comparison with the spacecraft reference axes shown in Fig.~\ref{fig:EPD-FOVs}. This projection serves as a geometric visualization tool and although the true angular responses of neighbouring STEP pixels overlap, the map is sufficient to identify where the signal is predominantly detected relative to both STEP and EPT-Sun FOVs and whether the central direction of the beam changes.

Magnetic field measurements from MAG are expressed in the spacecraft reference system and converted to elevation and azimuth angles. For each event, the magnetic field is averaged over time, using the same resampling as applied to the electron flux when determining its peak intensities to plot the associated in-situ spectrum. For each SEE event where EPT–Sun measurements are combined with STEP observations, the resulting field direction at the peak time for the STEP energy channels included in the spectrum fitting are then overlaid on the left panels of Figs. \ref{fig:MAG-STEP+EPT_FOVs_20221224} to \ref{fig:MAG-STEP+EPT_FOVs_20221112}, while the right panels assess the resulting field direction at the peak time for each EPT-Sun energy channel included in the fittings. We note that while only STEP energies above 20 keV are included in the spectrum for the event on 12 November 2022, as shown in panel (e) of Fig. \ref{all-is-spectra}, all STEP energies with significant statistics (above the mean background+3$\sigma$) remain represented in Fig. \ref{fig:MAG-STEP+EPT_FOVs_20221112}. 

This diagnostic allows us to identify cases where the magnetic field points toward outer STEP pixels or partially outside the common STEP/EPT–Sun FOV at the time of the peak flux. In such cases, individual energy channels may not sample the bulk of the electron beam, potentially introducing artificial spectral features or shifting apparent spectral breaks. The maps presented in this appendix Figs. \ref{fig:MAG-STEP+EPT_FOVs_20221224} to \ref{fig:MAG-STEP+EPT_FOVs_20221112}, therefore, provide a consistency check for the combined STEP+EPT–Sun spectra and help distinguish physical spectral features from effects related to viewing geometry and magnetic connectivity.

In the 24~December~2022 event, which is fitted with a double broken power law as shown in panel (c) of Fig.~\ref{fig:enu-spec-is}, the MAG data in Fig.~\ref{fig:MAG-STEP+EPT_FOVs_20221224} show that the magnetic-field direction associated with the peak times of all STEP and EPT–Sun energy channels used in the fitting consistently points in the same direction within the shared FOV. This indicates that both instruments sample the main electron beam across the entire fitted energy range, supporting the interpretation that the two detected spectral transitions are of physical origin rather than artifacts.

\begin{figure*}[ht!]
    \centering
        \includegraphics[width=\linewidth]{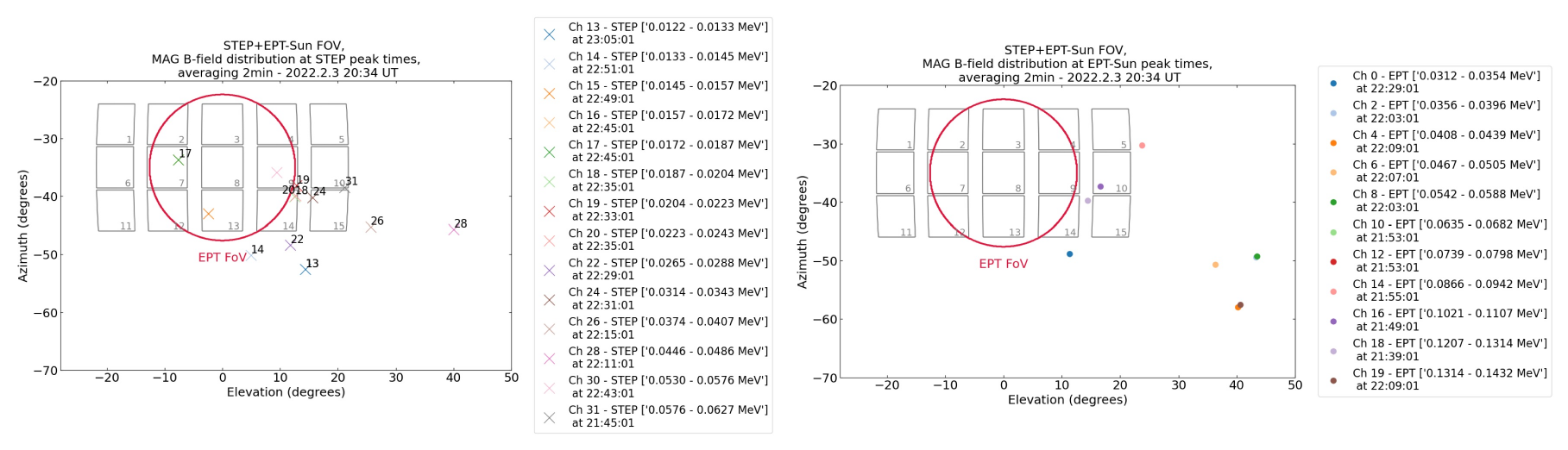}
        %\caption{}
    \caption{Sky-projected representation of the FOV of STEP sensor, with its pixels mapped into elevation-azimuth coordinates, for the 3 February 2022 SEE event. Panels as in Fig. \ref{fig:MAG-STEP+EPT_FOVs_20221224}.}
    \label{fig:MAG-STEP+EPT_FOVs_20220203}
\end{figure*}

In the 3~February~2022 event, the magnetic-field directions corresponding to the STEP channels included in the fitting, in panel (a) of Fig~\ref{fig:enu-spec-is}, are more scattered. This is illustrated by Fig.~\ref{fig:MAG-STEP+EPT_FOVs_20220203}: several energies do not point to STEP pixels lying within the EPT–Sun FOV, and a clear change in field orientation occurs near the $\sim$20~keV range where the spectral break is observed. This suggests that the break in this event is more likely to be artificial, with peak fluxes arising from pixels outside the common FOV. For clarity, the outward-pointing STEP channels, and more generally those in which we observe the change in field orientation near the $\sim$20~keV range, are explicitly labelled in left panel of Fig.~\ref{fig:MAG-STEP+EPT_FOVs_20220203}.

In the case of the 12~November~2022 event, a clear change in the magnetic-field direction appears around $\sim$28.8~keV, where the peak-time field for the highest STEP energies points outside the lower edge of both the STEP and EPT–Sun FOVs. These are explicitly labelled in left panel of Fig.~\ref{fig:MAG-STEP+EPT_FOVs_20221112}. However, no spectral break is detected in this range, as shown in panel (e) of Fig.~\ref{all-is-spectra}. This suggests that in this case, these energy-dependent variations in magnetic connectivity, does not seem to strongly influence the spectral shape, which is fitted by a single power-law. One possible reason is that the particle beam may be relatively wide compared to the STEP and EPT–Sun fields of view for this particular event. This should be considered when interpreting combined spectra. 

\begin{figure*}[ht!]
    \centering
        \includegraphics[width=\linewidth]{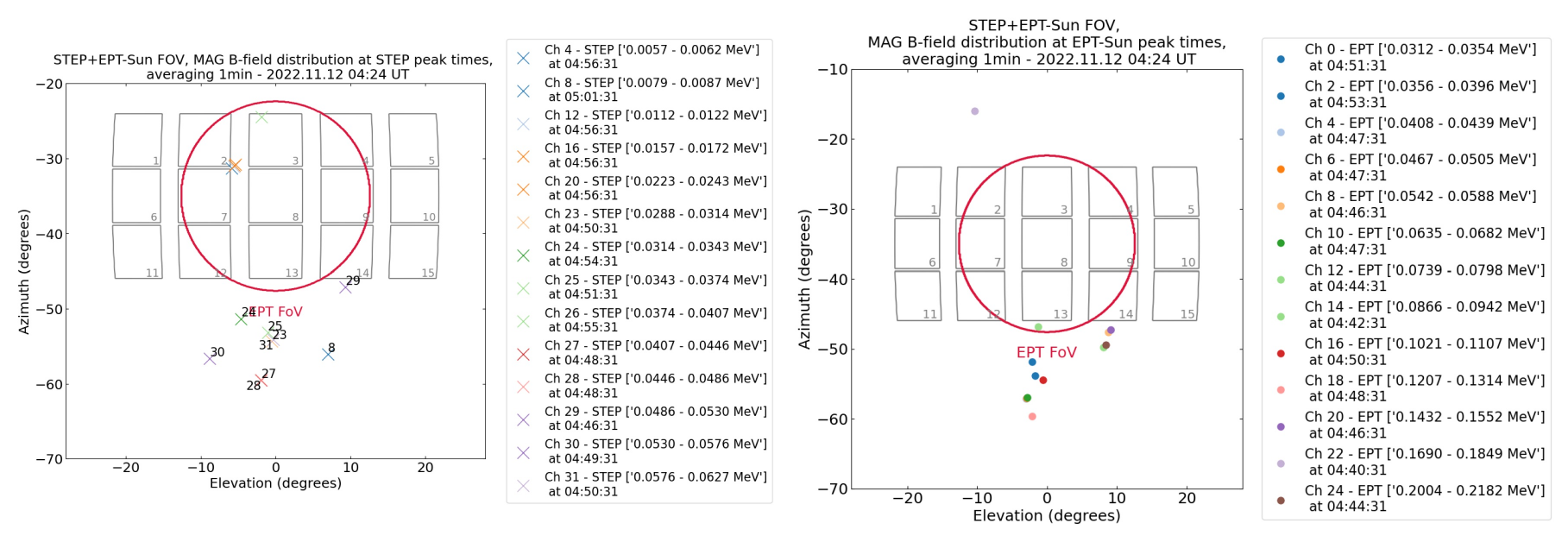}
        %\caption{}
        \label{fig:MAG-STEP+EPT_FOVs_20221112}
    \caption{Sky-projected representation of the FOV of STEP sensor, with its pixels mapped into elevation-azimuth coordinates, for the 12 November 2022 SEE event. Panels as in Fig. \ref{fig:MAG-STEP+EPT_FOVs_20221224}.}
    \label{fig:MAG-STEP+EPT_FOVs_20221112}
\end{figure*}

%\clearpage
\onecolumn

\clearpage

\end{appendix}

\end{document}